\documentclass[10pt,twocolumn]{IEEEtran}

\usepackage{graphicx}
\usepackage{amssymb}
\usepackage{amsbsy}
\usepackage{amsmath}

\usepackage{cite}
\usepackage{stfloats}
\usepackage{url}

\usepackage{enumerate}
\usepackage{color}
\usepackage{placeins}
\usepackage{float}
\usepackage{tabularx,colortbl}
\usepackage{times,amsmath,epsfig}
\usepackage{xspace,latexsym,syntonly}
\usepackage{amssymb}
\usepackage{amsfonts}
\usepackage{textcomp}
\usepackage{caption}
\usepackage{subcaption}
\usepackage{amsbsy}
\usepackage{cite}
\usepackage{mathtools, cuted}
\usepackage{lipsum, color}

\newcommand{\E}{\ensuremath{\hbox{\textbf{E}}}}

\newtheorem{theorem}{Theorem}
\newtheorem{lemma}{Lemma}

\newtheorem{definition}{Definition}
\newtheorem{corollary}{Corollary}

\newcommand{\beq}{\begin{equation}}
\newcommand{\eeq}{\end{equation}}
\newcommand{\bea}{\begin{array}}
\newcommand{\ena}{\end{array}}
\newcommand{\bds}{\begin {itemize}}
\newcommand{\eds}{\end {itemize}}
\newcommand{\bdf}{\begin{definition}}
\newcommand{\blm}{\begin{lemma}}
\newcommand{\edf}{\end{definition}}
\newcommand{\elm}{\end{lemma}}
\newcommand{\bthm}{\begin{theorem}}
\newcommand{\ethm}{\end{theorem}}
\newcommand{\bprp}{\begin{prop}}
\newcommand{\eprp}{\end{prop}}
\newcommand{\bcl}{\begin{claim}}
\newcommand{\ecl}{\end{claim}}
\newcommand{\bcr}{\begin{coro}}
\newcommand{\ecr}{\end{coro}}
\newcommand{\bquest}{\begin{question}}
\newcommand{\equest}{\end{question}}

\newcommand{\larrow}{{\larrow}}

\def\urltilda{\kern -.15em\lower .7ex\hbox{\~{}}\kern .04em}

\begin{document}\title{Active Anomaly Detection in Heterogeneous Processes}
\author{Boshuang Huang, Kobi Cohen, Qing Zhao
\thanks{Boshuang Huang and Qing Zhao are with the School of Electrical and Computer Engineering, Cornell University, Ithaca, NY, 14853, USA. Email: \{bh467, qz16\}@cornell.edu. Kobi Cohen is with the Department of Electrical and Computer Engineering, Ben-Gurion University of the Negev, Beer-Sheva 84105, Israel. Email: yakovsec@bgu.ac.il.}
\thanks{This work was supported by the National Science Foundation under Grant CCF-1815559 and by the Army Research Office under Grant W911NF-17-1-0464. The work of Kobi Cohen was supported by the Cyber Security Research Center at Ben-Gurion University of the Negev, and the U.S.-Israel Binational Science Foundation (BSF) under grant 2017723.}
}
\date{}
\maketitle
\begin{abstract}
\label{sec:abstract}
An active inference problem of detecting anomalies among heterogeneous processes is considered. At each time, a subset of processes can be probed. The objective is to design a sequential probing strategy that dynamically determines which processes to observe at each time and when to terminate the search so that the expected detection time is minimized under a constraint on the probability of misclassifying any process. This problem falls into the general setting of sequential design of experiments pioneered by Chernoff in 1959, in which a randomized strategy, referred to as the Chernoff test, was proposed and shown to be asymptotically optimal as the error probability approaches zero. For the problem considered in this paper, a low-complexity deterministic test is shown to enjoy the same asymptotic optimality while offering significantly better performance in the finite regime and faster convergence to the optimal rate function, especially when the number of processes is large. The computational complexity of the proposed test is also of a significantly lower order. 
\end{abstract}
%
\def\keywords{\vspace{.5em}
{\bfseries\textit{Index Terms}---\,\relax%
}}
\def\endkeywords{\par}
\keywords
Active hypothesis testing, sequential design of experiments, anomaly detection, dynamic search, target whereabout.

\section{Introduction}
\label{sec:intro}
We consider the problem of detecting an anomalous process among $M$ heterogeneous processes. Borrowing terminologies from target search, we refer to these processes as cells and the anomalous process as the target which can locate in any of the $M$ cells. At each time, $K$ ($1\le K<M$) cells can be probed simultaneously to search for the target. Each search of cell $i$ generates a noisy observation drawn i.i.d. over time from two different distributions $f_i$ and $g_i$, depending on whether the target is absent or present. The objective is to design a sequential search strategy that dynamically determines which cells to probe at each time and when to terminate the search so that the expected detection time is minimized under a constraint on the probability of declaring a wrong location of the target.

The above problem is prototypical of searching for rare events  in a large number of data streams or a large system. The rare events could be opportunities (e.g., financial trading opportunities or transmission opportunities
in dynamic spectrum access~\cite{zhao2007survey}), unusual activities in surveillance feedings, frauds in financial transactions, attacks and intrusions in communication and computer networks, anomalies in infrastructures (such as bridges, buildings, and the power grid) that may indicate catastrophes. Depending on the application, a cell may refer to an autonomous data stream with a continuous data flow or a system component that only generates data when probed.

\subsection{Main Results}
\label{ssec:main}
The anomaly detection problem considered in this paper is a special case of active hypothesis testing originated from Chernoff's seminal work on sequential design of experiments in 1959~\cite{chernoff1959sequential}. Compared with the classic passive sequential hypothesis testing pioneered by Wald~\cite{wald1947sequential}, where the observation model under each hypothesis is predetermined, active hypothesis testing has a control aspect that allows the decision maker to choose the experiment to be conducted at each time. Different experiments generate observations from different distributions under each hypothesis. Intuitively, as more observations are gathered, the decision maker becomes more certain about the true hypothesis, which in turn leads to better choices of experiments.

In~\cite{chernoff1959sequential}, Chernoff proposed a \emph{randomized} strategy, referred to as the Chernoff test, and established its asymptotic (as the error probability diminishes) optimality\footnote{The asymptotic optimality of the Chernoff test was shown under the assumption that the hypotheses are distinguishable under every experiment.}. This randomized test chooses, at each time, a probability distribution that governs the selection of the experiment to be carried out at this time. This distribution is obtained by solving a minimax problem so that the next observation generated under the random action can best differentiate the current maximum likelihood estimate of the true hypothesis (using all past observations) from its closest alternative, where the closeness is measured by the Kullback-Liebler (KL) divergence. Due to the complexity in solving this minimax problem at each time, the Chernoff test can be expensive to compute and cumbersome to implement, especially when the number of hypotheses or the number of experiments is large. 

It is not difficult to see that the problem at hand is a special case of the general active hypothesis testing problem. Specifically, the available experiments are in the form of different subsets of $K$ cells to probe, and the number of experiments is $\binom{M}{K}$. Under each hypothesis that cell $m$ $(m = 1,\ldots,M)$ is the target, the distribution of the next observation (a vector of dimension $K$) depends on which $K$ cells are chosen.  The Chernoff test thus directly applies. Unfortunately, with the large number of hypotheses and the large number of experiments, it can be computationally prohibitive to obtain the Chernoff test. 

In this paper, we show that the anomaly detection problem considered here exhibits sufficient structures to admit a low-complexity \emph{deterministic} policy with strong performance. In particular, we develop a deterministic test that \emph{explicitly} specifies which $K$ cells to search at each given time and show that this test enjoys the same asymptotic optimality as the Chernoff test\footnote{The asymptotic optimality of the proposed test holds for all but at most three singular values of $K$ (see Theorem~\ref{th:optimality_policy3}).}. Furthermore, extensive simulation examples have demonstrated significant performance gain over the Chernoff test in the finite regime and faster convergence to the optimal rate function, especially when $M$ is large. In contrast to the Chernoff test, the proposed test requires little offline or online computation. The test can also be extended to cases with multiple targets as discussed in Section~\ref{sec:multitarget}. Its asymptotic optimality is preserved for $K=1$.

Often, when a solution is simpler, establishing its optimality becomes harder. This is indeed the case here. In Chernoff test, since the distribution of the random action depends only on the current maximum likelihood estimate of the underlying hypothesis which becomes time-invariant after an initial phase with a bounded duration, the stochastic behaviors of the test statistics, namely, the log-likelihood ratios (LLRs),  are independent over time. In contrast, the deterministic actions under the proposed policy result in strong time and spacial (across processes) dependencies in the dynamic evolutions of the LLRs. Establishing the asymptotic optimality becomes much more involved.

\subsection{Related Work}

Chernoff's pioneering work on sequential design of experiments
focuses on binary composite hypothesis testing~\cite{chernoff1959sequential}. Variations and
extensions have been studied
in~\cite{bessler1960theory,nitinawarat2012controlled,nitinawarat2013controlled,nitinawarat2015controlled,
naghshvar2013active,naghshvar2013sequentiality}, where the problem was referred to as controlled
sensing for hypothesis testing in~\cite{nitinawarat2012controlled,nitinawarat2013controlled,nitinawarat2015controlled} and active hypothesis testing in~\cite{naghshvar2013active,naghshvar2013sequentiality}. As variants of the Chernoff test, the tests developed in~\cite{bessler1960theory,nitinawarat2012controlled,nitinawarat2013controlled,nitinawarat2015controlled,
naghshvar2013active,naghshvar2013sequentiality} are all randomized tests. 

There is an extensive literature on dynamic search and target whereabout problems under various scenarios. We discuss here existing studies within the sequential inference setting, which is the most relevant to this work. Two models on prior information about the targets have been considered in the literature: the exclusive model which assumes a fixed number of targets and the independent model which assumes each cell may contain a target with a given prior probability independent of other cells. These two models were juxtaposed in~\cite{castanon1995optimal,cohen2014optimal} under different objective functions. The studies in~\cite{vaidhiyan2015learning,leahy2016always,heydari2016quickest,cohen2015active,zigangirov1966problem} focus on the exclusive model. In particular, homogeneous Poisson point processes with unknown rates was investigated and an asymptotically optimal randomized test was developed in \cite{vaidhiyan2015learning}. In~\cite{leahy2016always}, the problem of tracking a target that moves as a Markov Chain in a finite discrete environment is studied and a search strategy that provides the most confident estimate is developed. The studies in \cite{tajer2013quick,cohen2015asymptotically,fellouris2017multistream,lai2011quickest} focus on the independent model. The problem of searching among Gaussian signals with rare mean and variance values was studied and an adaptive group sampling strategy was developed in \cite{tajer2013quick}. In \cite{cohen2015asymptotically}, the problem of quickly detecting anomalous components under the objective of minimizing system-wide cost incurred by all anomalous components was studied. In \cite{fellouris2017multistream}, an important case of multichannel sequential change detection is studied and an asymptotic framework in which the number of sensors tends to infinity was proposed.

Asymptotically optimal search policies over homogeneous processes were established in~\cite{nitinawarat2015universal} under a non-parametric setting with finite discrete distributions and in~\cite{hemo2016} under a parametric composite hypothesis setting with continuous distributions.  The objective of minimizing operational cost as opposed to detection delay led to a different problem from the one considered in this paper. Other related work on quickest search over multiple processes under various models and formulations includes~\cite{castanon1995optimal, lai2011quickest, malloy2012quickest, heydari2016quickest} and references therein. Sequential spectrum sensing  within both the passive and active hypothesis testing frameworks has also received extensive attention in the application domain of cognitive radio networks (see, for example, \cite{pei2011energy, caromi2013fast, ferrari2017utility, egan2017fast} and references therein). The readers are also referred to~\cite{tajer2014outlying} for a comprehensive survey on the problem of detecting outlying sequences.

A prior study by Cohen and Zhao considered the problem for  homogeneous processes (i.e., $f_i\equiv f$ and $g_i\equiv g$)~\cite{cohen2015active}. This work builds upon this prior work and addresses the problem in heterogeneous systems where the absence distribution $f_i$ and the presence distribution $g_i$  are different across processes. Allowing heterogeneity significantly complicates the design of the test and the analysis of asymptotic optimality. Since each process has different observation distributions, the rate at which the state of a cell can be inferred is different across processes. To achieve asymptotic optimality, the decision maker must carefully balance the search time among the observed processes, which makes both the algorithm design and the performance analysis much more involved under the heterogeneous case. Specifically, in terms of algorithm design, when dealing with homogeneous processes, the search strategy is often static in nature~\cite{castanon1995optimal,vaidhiyan2015learning,nitinawarat2015universal,cohen2015active}. In contrast, the asymptotically optimal search strategy developed here for heterogeneous processes dynamically changes based on the current belief about the location of the target. In terms of performance analysis, when dealing with homogeneous processes, the resulting rate function (which is inversely proportional to the search time) always obeys a certain averaging over the KL divergences between normal and abnormal distributions of all processes. This observation follows from the fact that the decision maker completes gathering the required information from all the processes at approximately the same time due to the homogeneity. In contrast, when searching over heterogeneous processes, the overall rate function does not always obey a simple averaging across the KL divergences of all processes. In Section~\ref{sec:performance}, we show that the search time can be analyzed by considering two separate scenarios, referred to as the balanced and the unbalanced cases. The balanced case holds when a judicious allocation of probing resources can ensure the information gathering from all the processes be completed at approximately the same time, in which case the rate function is a weighted average among the heterogeneous processes. The unbalanced case occurs when there is a process with a sufficiently small KL divergence that it dominates the overall rate function of the search. This case is unique to the heterogeneous processes considered here and needs to be addressed with new analytical techniques. 

Besides the active inference approach to anomaly detection considered in this paper, there is a growing body of literature on various approaches to the general problem of anomaly detection. We refer the readers to~\cite{chandola2009anomaly,bhuyan2014network} for comprehensive surveys on this topic.

\section{Problem Formulation}
\label{sec:problem}

We consider the problem of detecting a single target located in one of $M$ cells. 
If the target is in cell $m$, we say that hypothesis $H_m$ is true. The \emph{a priori} probability that $H_m$ is true is denoted by $\pi_m$, where $\sum_{m=1}^M \pi_m=1$. To avoid trivial solutions, it is assumed that $0<\pi_{m}<1$ for all $m$.

When cell $m$ is observed at time~$n$, an observation $y_m(n)$ is drawn,  independent of previous observations. If cell $m$ contains a target, $y_m(n)$ follows distribution $g_m(y)$. Otherwise, $y_m(n)$ follows distribution $f_m(y)$. Let $\mathbf{P}_{m}$ be the probability measure under hypothesis $H_m$ and $\E_{m}$ the operator of expectation with respect to the measure $\mathbf{P}_{m}$.

An active search strategy $\Gamma$ consists of a stopping rule $\tau$ governing when to terminate the search, a decision rule $\delta$ for determining the location of the target at the time of stopping, and a sequence of selection rules $\{\phi(n)\}_{n\ge 1}$ governing which $K$ cells to probed at each time $n$. 
 Let $\mathbf{y}(n)$ be the set of all cell selections and observations up to time~$n$. A deterministic selection rule $\phi(n)$ at time~$n$ is a mapping from $\mathbf{y}(n-1)$ to $\left\{1, 2, \ldots, M\right\}^K$. A randomized selection rule $\phi(n)$ is a mapping from $\mathbf{y}(n-1)$ to probability mass functions over $\left\{1, 2, \ldots, M\right\}^K$.

We adopt a Bayesian approach as in Chernoff's original study~\cite{chernoff1959sequential} by assigning a cost of $c$ for each observation and a loss of $1$ for a wrong declaration. Note that $c$ represents the ratio of the sampling cost to the cost of wrong detections. The Bayes risk under strategy $\Gamma$ when hypothesis $H_m$ is true is given by:
\beq
\label{eq:Bayes_risk_m}
\displaystyle R_{m}(\Gamma)\triangleq\alpha_{m}(\Gamma)+c\E_{m}(\tau|\Gamma),
\eeq
where $\alpha_{m}(\Gamma)=\mathbf{P}_{m}(\delta \neq m |\Gamma)$ is the probability of declaring $\delta\neq {m}$ under $H_m$ and $\E_{m}(\tau|\Gamma)$ is the detection delay under $H_m$. The average Bayes risk is given by:
\beq
\label{eq:Bayes_risk}
\displaystyle R(\Gamma)=\sum_{m=1}^{M}\pi_{m} R_{m}(\Gamma)=P_e(\Gamma)+c\E(\tau|\Gamma),
\eeq
where $P_e(\Gamma)$ and $\E(\tau|\Gamma)$ are the error probability and detection delay averaged under the given prior $\{\pi_m\}$. The objective is to find a strategy $\Gamma$ that minimizes the Bayes risk $R(\Gamma)$:
\beq\label{eq:Bayes_formulation1}
\displaystyle\inf_{\Gamma} \;\; R(\Gamma) .
\eeq

A strategy $\Gamma^*$ is \emph{asymptotically optimal} if
\beq
\lim_{c\rightarrow 0} \;\frac{R(\Gamma^*)}{\inf_{\Gamma} R(\Gamma)}=1,
\eeq
which is denoted as 
\beq
R(\Gamma^*) \sim \inf_{\Gamma} R(\Gamma).
\eeq

\section{The Deterministic DGF\MakeLowercase{i} Policy}
\label{sec:DGFi}

In this section we propose a deterministic policy, referred to as the DGFi policy, indicating the key quantities $\{D(g_i||f_i),\,D(f_i||g_i)\}_{i=1}^M$ that govern the selection rule of the proposed policy. 
\subsection{DGFi under Single-Cell Probing}
\label{subsec:simple}

We first consider the case of $K=1$. Let $\mathbf{1}_m(n)$ be the indicator function, where $\mathbf{1}_m(n)=1$ if cell $m$ is observed at time~$n$, and $\mathbf{1}_m(n)=0$ otherwise. This indicator function clearly depends on the selection rule, which we omit in the notation for simplicity. Let
\beq
\label{eq:LLR}
\displaystyle \ell_m(n)\triangleq\log \frac{g_m(y_m(n))}{f_m(y_m(n))} \;,
\eeq
and
\beq
\label{eq:sum_LLR}
\displaystyle S_m(n)\triangleq\sum_{t=1}^{n}{\ell_m(t)\mathbf{1}_m(t)}
\eeq
be the LLR and the observed sum LLRs of cell $m$ at time~$n$, respectively.  
Let $D(g||f)$ denote the KL divergence between two distributions $g$ and $f$ which is given by\footnote{We assume that $g_i$ is absolutely continuous with respect to $f_i$ ($i=1,\ldots, M$) and vise versa, which ensures that all KL divergences are finite.}
\beq
\label{eq:defdfg}
D(g||f)\triangleq \int_{-\infty}^{\infty} \log\frac{g(x)}{f(x)}g(x) \;dx.
\eeq

\begin{figure}[!h]
\centering
\includegraphics[width=3in]{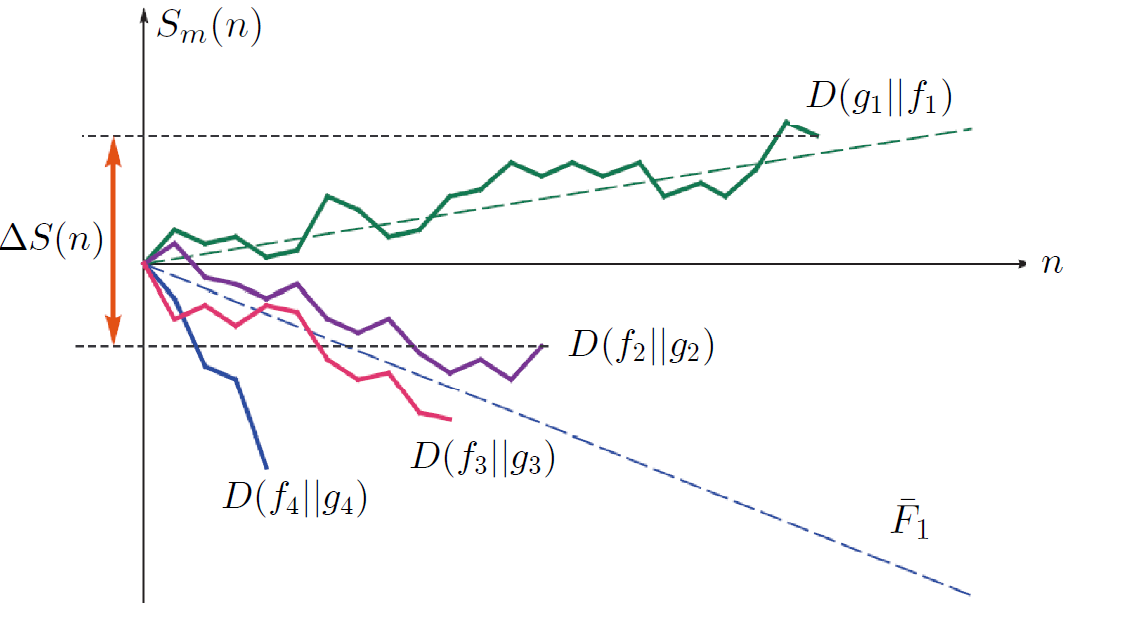}
\caption{Typical sample paths of sum LLRs.}
\label{fig_DGFintui}
\end{figure}
Illustrated in Fig.~\ref{fig_DGFintui} are typical sample paths of the sum LLRs of $M=4$ cells, where, without loss of generality, we assume that cell 1 is the target. Note that the sum LLR of cell 1 is a random walk with a positive expected increment $D(g_1||f_1)$, whereas the sum LLR of cell $m$ ($m=2,3,4$) is a random walk with a negative expected increment $-D(f_m||g_m)$. Thus, when the gap between the largest sum LLR and the second largest sum LLR is sufficiently large, we can declare with a sufficient accuracy that the cell with the largest sum LLR is the target. This is the intuition behind the stopping rule and the decision rule under DGFi. Specifically, we define $m^{(i)}(n)$ as the index of the cell with the $i^{th}$ largest observed sum LLRs at time~$n$. Let
\beq
\label{eq:Delta_S}
\Delta S(n)\triangleq S_{m^{(1)}(n)}(n)-S_{m^{(2)}(n)}(n)
\eeq
denote the difference between the largest and the second largest observed sum LLRs at time~$n$. The stopping rule and the decision rule under the DGFi policy are given by:
\beq
\label{eq:stopping_policy1}
\bea{l}
\displaystyle \tau= \inf \left\{n \; : \; \Delta S(n)\geq -\log c\right\}\;,
\ena
\eeq
and
\beq
\label{eq:decision_policy1}
\displaystyle\delta= m^{(1)}(\tau) \;.
\eeq

We now specify the selection rule of the DGFi policy. The intuition behind the selection rule is to select a cell from which the observation can increase $\Delta S(n)$ at the fastest rate. The selection rule is thus given by comparing the rate at which $S_{m^{(1)}(n)}(n)$ increases with the rate at which $S_{m^{(2)}(n)}(n)$ decreases. If $S_{m^{(1)}(n)}(n)$ is expected to increase faster than $S_{m^{(2)}(n)}(n)$ decreases, cell $m^{(1)}(n)$ is chosen. Otherwise, cell $m^{(2)}(n)$ is chosen. This leads to the following selection rule: 
\begin{equation}
\label{eq:selection}
\phi(n) = \left\{\begin{array}{ll}
m^{(1)} (n), & \mbox{if } D(g_{m^{(1)}(n)}||f_{m^{(1)}(n)})\geq \bar F_{m^{(1)}(n)}\\
m^{(2)}(n), & \mbox{otherwise}\\
\end{array}
\right.,
\end{equation}
where
\beq
\label{eq:fm}
\bea{l}
\displaystyle \bar F_m \triangleq \frac{1}{\sum_{j\neq m}\frac {1}{D(f_j||g_j)}}.
\ena
\eeq

 The selection rule in \eqref{eq:selection} can be intuitively understood by noticing that $D(g_{m^{(1)}(n)}||f_{m^{(1)}(n)})$ is the asymptotic increasing rate of $S_{m^{(1)}}(n)$ when cell $m^{(1)}$ is probed at each time. This is due to the fact that $m^{(1)}(n)$ is the true target after an initial phase (defined by the last passage time that $m^{(1)}(n)$ is an empty cell) which can be shown to have a bounded expected duration. Similarly, even though much more involved to prove, $\bar F_{m^{(1)}(n)}$ is the asymptotic rate at which $S_{m^{(2)}(n)}(n)$ decreases when cell $m^{(2)}(n)$ is probed at each time. To see the expression of $\bar F_m$ for any $m$ as given in \eqref{eq:fm}, consider the following analogy. Consider $M-1$ cars being driven by a single driver from $0$ to $-\infty$. Car $j$ ($j=1,\ldots, M$, $j\neq m$) has a constant speed of $D(f_j||g_j)$. At each time, the car closest to the origin is chosen by the driver and driven by one unit of time. We are interested in the average moving speed of the position of the closest car to the origin. It is not difficult to see that it is given by $\bar F_m$ in \eqref{eq:fm}. This analogy, concerned with deterministic processes, only serves as an intuitive explanation for the expression of $\bar F_m$. As detailed in Sec.~\ref{sec:performance}, proving $\bar F_{m^{(1)}(n)}$ to be the asymptotic decreasing rate of $S_{m^{(2)}(n)}(n)$ requires analyzing the trajectories of the $M$ sum LLRs $\{S_m(n)\}_{m=1}^M$, which are stochastic processes with complex dependencies both in time and across processes.

\subsection{DGFi under Multiple Simultaneous Observations}
\label{subsec:extK}
Now we consider the case of $K>1$. The stopping rule and the decision rule remains the same as given in~\eqref{eq:stopping_policy1},~\eqref{eq:decision_policy1}, whereas the selection rule requires a significant modification.  The main reason is that when $K$ cells can be observed simultaneously, the asymptotic increasing rate of $S_{m^{(1)}(n)}(n)$ and the asymptotic decreasing rate of $S_{m^{(2)}(n)}(n)$ are much more involved to analyze.

 The selection rule $\phi(n)$, at each time~$n$, chooses either the $K$ cells with the top $K$ largest sum LLRs or those with the second to the $(K+1)^{th}$ largest sums LLRs as in (14) where 
 
\setcounter{equation}{14}
\beq
\label{eq:fmk}
\bea{l}
\displaystyle F_{m}(\kappa)\triangleq \min\{\kappa  \bar F_m,~\min_{j\neq m} D(f_j||g_j)\}.
\ena
\eeq

Note that~\eqref{eq:fmk} reduces to~\eqref{eq:fm} at $K=1$ (i.e., $F_m(1)=\bar{F}_m$), in which case the minimum is always attained at the first term. Similar to the case with $K=1$, the intuition behind the selection rule is to select $K$ cells from which the observations increase $\Delta S(n)$ at the fastest rate.  Specifically, $F_{m^{(1)}(n)}(K)$ is the asymptotic decreasing rate of $S_{m^{(2)}(n)}(n)$ when $K$ cells with the second largest to the $(K+1)$th largest sum LLRs are probed each time. When the cell with the top $K$ largest sum LLRs are probed each time, the asymptotic increasing rate of $\Delta S(n)$ is $D(g_{m^{(1)}(n)}||f_{m^{(1)}(n)})+F_{m^{(1)}(n)}(K-1)$, where $D(g_{m^{(1)}(n)}||f_{m^{(1)}(n)})$ is the asymptotic increasing rate of $S_{m^{(1)}(n)}(n)$ and $F_{m^{(1)}(n)}(K-1)$ is the asymptotic decreasing rate of $S_{m^{(2)}(n)}(n)$ with $K-1$ drivers. It is easy to see that when $K=1$, the policy reduces to the one described in section~\ref{subsec:simple}.

\begin{figure}[h]
\centering
\includegraphics[scale=0.3]{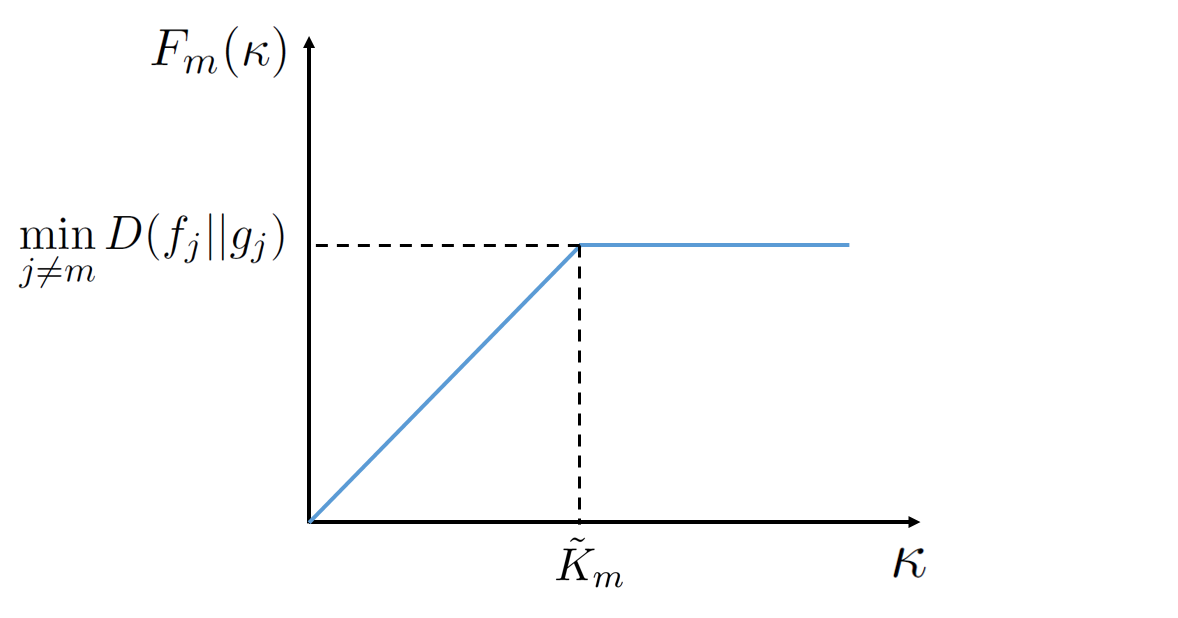} 
\caption{The piecewise linear property of $F_m(\kappa).$}
\label{fig:fmk}
\end{figure}

The behavior of $F_m(\kappa)$ as a function of $\kappa$ (extending $\kappa$ to all positive real values) is crucial in understanding and analyzing the asymptotic optimality of DGFi for $K>1$. It is easy to see that the first term in the right hand of \eqref{eq:fmk} is a linearly increasing function of $\kappa$ and the second term is a constant. This readily leads to the piecewise linear property of $F_m(\kappa)$ as illustrated in Fig.~\ref{fig:fmk}. Let $\tilde{K}_m$ denote the switching point between the increasing and constant regions, we have
\vspace{-1.5cm}
\begin{strip}
\hrulefill
\beq
\label{eq:dgfk}
\bea{l}
\displaystyle \phi(n)=
\begin{cases} \left(m^{(1)}(n), m^{(2)}(n), ..., m^{(K)}(n)\right) 
                                          &      \mbox{if $D(g_{m^{(1)}(n)}||f_{m^{(1)}(n)})+F_{m^{(1)}(n)}(K-1)\ge F_{m^{(1)}(n)}(K)$}  \\
              \left(m^{(2)}(n), m^{(3)}(n), ..., m^{(K+1)}(n)\right) &\mbox{otherwise}
\end{cases}
\tag{14}
\ena
\eeq
\end{strip}

\begin{equation}
\label{eq:kmtilde}
\tilde{K}_m =\frac{\min_{j\neq m}D(f_j||g_j)}{\bar F_m} = \sum_{j\neq m}\frac{\min_{j\neq m}D(f_j||g_j)}{D(f_j||g_j)}.
\end{equation}

The constant value of $F_m(\kappa)$ for $\kappa \ge \tilde{K}_m$ can be explained with the same car analogy. This constant value $\min_{j\neq m} D(f_j||g_j)$ is the speed of the slowest car among the $M-1$ cars (excluding the $m$th car). When the speed of the slowest car is sufficiently small, this car always lags behind even with a dedicated driver. This car becomes the bottleneck that caps the value of $F_m(\kappa)$ even when the number $\kappa$ of drivers increases (note that each car can at most have one driver assigned). We refer to this case as the unbalanced case, which presents the most challenge in proving the asymptotic optimality of DGFi. The linearly increasing region of $\kappa<\tilde{K}_m$ is referred to as the balanced case, where $F_m(\kappa)$ is a weighted average among the $M-1$ cars.

\section{Performance Analysis}
\label{sec:performance}
In this section, we establish the asymptotic optimality of the DGFi policy. While the intuitive exposition of DGFi given in Sec.~\ref{sec:DGFi} may make its asymptotic optimality seem expected, constructing a proof is much more involved. In particular, bounding the detection time of DGFi requires analyzing the trajectories of the $M$ stochastic processes $\{S_m(n)\}_{m=1}^M$ which exhibit complex dependencies both over time and across processes as induced by the deterministic selection rule.

The asymptotic optimality of DGFi is established by comparing its Bayes risk (given in Theorem~\ref{th:optimality_policy1}) with a lower bound on achievable Bayes risk (given in Theorem~\ref{th:optimality_policy2}). We first analyze the rate function of DGFi. Define 
\beq
\bea{l}
\label{eq:I_m}
\displaystyle I_m (\Gamma_{\mbox{\footnotesize DGFi}})\triangleq \max\{D(g_m||f_m)+F_m(K-1), F_m(K)\},
\ena
\eeq
which is the increasing rate of $\Delta S(n)$ under hypothesis $H_m$ when DGFi is employed. For a given \emph{a priori} distribution $\{\pi_m\}_{m=1}^M$, define
\beq
\label{eq:I}
\bea{l}
\displaystyle I (\Gamma_{\mbox{\footnotesize DGFi}}) \triangleq \frac 1{\sum_{m=1}^{M} \frac {\pi_{m}}{I_m (\Gamma_{\mbox{\scriptsize DGFi}})}}.
\ena
\eeq
As shown in Theorem~\ref{th:optimality_policy1} below, $I (\Gamma_{\mbox{\footnotesize DGFi}})$ is the rate function of the Bayes risk of the DGFi policy. 
\vspace{0.2cm}

\begin{theorem}
\label{th:optimality_policy1}
The Bayes risk $R(\Gamma_{\mbox{\footnotesize DGFi}})$ of the DGFi policy is given by
\beq
\label{eq:dgfi_performance}
R(\Gamma_{\mbox{\footnotesize DGFi}}) \;\sim\; \frac{-c\log c}{I (\Gamma_{\mbox{\footnotesize DGFi}})}.
\vspace{0.2cm}
\eeq
\end{theorem}

\begin{IEEEproof}
Here we provide a sketch of the proof.  The detailed proof can be found in Appendix A. First, we show that when $\Delta S(\tau)$ is large, the probability of error is small, i.e. $P_e=O(c)$. As a result, by the definition of the Bayes risk, it suffices to show that the detection time is upper bounded by $-\log c/I (\Gamma_{\mbox{\footnotesize DGFi}})$. By the definition of $I (\Gamma_{\mbox{\footnotesize DGFi}})$ in (\ref{eq:I}), it suffices to show that the detection time is upper bounded by $-\log c/I_m (\Gamma_{\mbox{\footnotesize DGFi}})$ under hypothesis $H_m$. Since the decision maker might not complete to gather the required information from all the cells at the same time, we carry out the analysis by treating the balanced and the unbalanced cases separately.

\end{IEEEproof}

Next we estabilsh a lower bound on the Bayes risk achievable by any policy. Define

\begin{eqnarray}
\label{eq:Imstar}
 I_m^*  & \triangleq& \max_{u\in[0,1]} uD(g_m||f_m)+F_m(K-u). \\
 I^*   & \triangleq  &\frac 1{\sum_{m=1}^{M} \frac {\pi_{m}}{I_{m}^*}}.
\end{eqnarray}

Using the same car analogy, we can interpret $I^*_m$ as the maximum increasing rate of $\Delta S(n)$ under hypothesis $H_m$ with an optimal allocation of $u^* \in [0,1]$ driver to the target car. Comparing with the rate of DGFi under $H_m$ in~\eqref{eq:I_m}, we see that the deterministic nature of DGFi forces the allocation of drivers to the target to be either $0$ or $1$.  As shown in Theorem~\ref{th:optimality_policy2} below, $I^*$ is an upper bound on the rate function for any policy. 

\begin{theorem}
\label{th:optimality_policy2}
Let $R(\Gamma)$ be the Bayes risk under  an arbitrary policy $\Gamma$. We have
\beq
\label{eq:upper_performance}
\inf_{\Gamma}\;{R(\Gamma)} \;\sim\;\frac{-c\log c}{I^*}
\vspace{0.2cm}
\eeq
\end{theorem}

\begin{IEEEproof}
The outline of the proof is as follows. We first prove that if the Bayes risk is sufficiently small under strategy $\Gamma$, i.e., $R(\Gamma)=O(-c\log c)$, the difference between the largest sum LLRs and the second largest sum LLRs must be sufficiently large when the test terminates, i.e. $\Delta S(\tau)=\Omega(-\log c)$. Otherwise, it is not possible to achieve a risk $O(-c\log c)$ due to a large error probability. We then show that in order to make $\Delta S(n)$ sufficiently large, the sample size must be large enough, i.e.,  $\E[\tau|\Gamma]\ge\frac{-\log c}{I^*}$. Since each sample costs $c$, the total risk will be lower bounded by $\frac{-c\log c}{I^*}$ as desired. The detailed proof can be found in Appendix B.

\end{IEEEproof}

Establishing the asymptotic optimality of DGFi rests on comparing its rate function $I (\Gamma_{\mbox{\footnotesize DGFi}})$ with the optimal rate function $I^*$. The key thus lies in analyzing the optimizer $u_m^*$ in the right hand of \eqref{eq:Imstar} and showing whether and when it assumes integer values of $0$ and $1$ as used in DGFi. This is established in Lemma~\ref{th:optimality_policy1} that leads to the following necessary and sufficient condition for the asymptotic optimality of DGFi.

\vspace{0.2cm}

\begin{theorem}
\label{th:optimality_policy3}
A necessary and sufficient condition for the asymptotic optimality of the DGFi policy is that, for each $m=1,\ldots,M$, at least one of the following three statements is true
\begin{enumerate}[(a)]
\item $D(g_m||f_m)\ge \bar F_m$.
\item  $K\le \tilde K_m$. 
\item $K\ge \tilde K_m+1$.

\end{enumerate}

\end{theorem}

\vspace{0.2cm}
\begin{IEEEproof}
We first establish the following lemma on the maximizer $u_m^*$ that attains $I_m^*$ given in~\eqref{eq:Imstar}. The proof of this lemma is in Appendix C.

\begin{lemma}
\label{lemma:umstar}
Define

\beq
u_m^*  \triangleq \arg\max_{u\in[0,1]}  uD(g_m||f_m)+F_m(K-u).
\eeq
 Then,
\beq
\label{eq:umstar}
u_m^*=
\begin{cases}
1, & \mbox{ if }D(g_m||f_m)\ge \bar F_m \\
\min\{\max\{K-\tilde K_m,0\},1\},& \mbox{ if } D(g_m||f_m)<\bar F_m
 
\end{cases}.
\eeq 
\end{lemma}

From~\eqref{eq:umstar} in Lemma 1, $u_m^*$ takes the integer value of $0$ or $1$ if and only if at least one of the Statements (a), (b), (c) is true. Theorem~\ref{th:optimality_policy3} thus follows.
\end{IEEEproof}

\vspace{0.2cm}

\begin{corollary}
The DGFi policy is asymptotically optimal except for at most three values of $K \in \{2,3, \ldots, M\}$ for every given problem instance specified by $\{M,\{D(g_i||f_i),D(f_i||g_i)\}_{i=1}^M\}$. 
\end{corollary}

\begin{IEEEproof}
From Theorem~\ref{th:optimality_policy3}, it is easy to see that for each $m$, there is only one possible $K=\lceil \tilde K_m\rceil$, which is the least integer greater than or equal to $\tilde K_m$, that makes $I_m (\Gamma_{\mbox{\footnotesize DGFi}})<I_m^*$. Let $j'=\arg\min_j D(f_j||g_j)$. Since there is only one possible $K=\lceil\tilde K_{j'}\rceil$ that makes $I_{j'} (\Gamma_{\mbox{\footnotesize DGFi}})<I_{j'}^*$, it remains to show that there are only two possible values of $K=\lceil \tilde K_m\rceil$ that makes $I_m (\Gamma_{\mbox{\footnotesize DGFi}})<I_m^*$ when $m\neq j'$. Let
$$
V\triangleq\sum_{j=1}^M\frac{D(f_{j'}||g_{j'})}{D(f_j||g_j)}.
$$
Since $0\le\frac{D(f_{j'}||g_{j'})}{D(f_m||g_m)}\le 1$, we have
$$
\tilde K_m = \sum_{j\neq m} \frac{\min_{j\neq m}D(f_j||g_j)}{D(f_j||g_j)}=V-\frac{D(f_{j'}||g_{j'})}{D(f_m||g_m)} \in [V-1,V]
$$
for all $m\neq j'$. This implies that $\lceil \tilde K_m \rceil (m\neq j')$ can only take two possible integers as desired.
\end{IEEEproof}

\vspace{0.3cm}

The above corollary also indicates that for $K=1$, the DGFi policy is always asymptotically optimal. This can be easily seen since Statement (b) always holds for $K=1$. To find those pathological values of $K$ for which DGFi is not asymptotically optimal, we can compute $\lceil \tilde K_m \rceil$ defined in~\eqref{eq:kmtilde} for each $m=1,2,\ldots,M$. Since for each $m$, $\lceil \tilde K_m \rceil$ only requires $O(M)$ number of multiplication and summation, the computational complexity of finding those pathological values is $O(M^2)$. 

\section{Extension to Detecting Multiple Targets}
\label{sec:multitarget}
In this section we extend the DGFi policy to the case with $L>1$ targets. The number of hypotheses in this case is $\binom{M}{L}$. We consider first $K=1$. The stopping rule and decision rule of DGFi for $L>1$ are given below, similar in principle to those for $L=1$ as described in Section~\ref{sec:DGFi}:

\beq
\label{eq:stopping_policyl}
\bea{l}
\displaystyle \tau= \inf \left\{n \; : \; \Delta S_L(n)\geq -\log c\right\},
\ena
\eeq
\beq
\label{eq:decision_policyl}
\displaystyle\delta= \{m^{(1)}(\tau), m^{(2)}(\tau), \ldots, m^{(L)}(\tau)\},
\eeq
where
\beq
\label{eq:Delta_Sl}
\Delta S_L(n)\triangleq S_{m^{(L)}(n)}(n)-S_{m^{(L+1)}(n)}(n)
\eeq
denotes the difference between the $L^{th}$ and the $(L+1)^{th}$ largest observed sum LLRs at time~$n$.

For the selection rule, define, for a given set $\mathcal D\subset \{1,2,\ldots, M\}$ with $|\mathcal D|=L$,
\beq
\label{FD1}
\bea{l}
\displaystyle \bar F_{\mathcal D}\triangleq \frac{1}{\sum_{j\notin \mathcal D}\frac {1}{D(f_j||g_j)}}.
\ena
\eeq
Similar to $\bar F_m$ defined in~\eqref{eq:fm}, $F_{\mathcal D}$ can be viewed as the asymptotic increasing rate of $\Delta S_L(n)$ when the $L$ targets are given by set $\mathcal D$ and we probe the cell with the $(L+1)^{th}$ largest sum LLR. We also define
\beq
\label{GD1}
\bea{l}
\displaystyle  \bar G_{\mathcal D} \triangleq \frac{1}{\sum_{j\in\mathcal D}\frac {1}{D(g_j||f_j)}},
\ena
\eeq
which can be viewed as the asymptotic increasing rate for $\Delta S_L(n)$ when we probe the cell with the $L^{th}$ largest sum LLR. 

The selection rule follows the same design principle of maximizing the asymptotic increasing rate of $\Delta S_L(n)$, and is given by
\begin{equation}
\label{eq:selectionL}
\phi(n) = \left\{\begin{array}{ll}
m^{(L)} (n), & \mbox{if } \bar G_{\mathcal D(n)}\geq \bar F_{\mathcal D(n)}\\
m^{(L+1)}(n), & \mbox{otherwise}\\
\end{array}
\right.,
\end{equation}
where
\beq
\mathcal D(n)=\{m^{(1)}(n),m^{(2)}(n),\ldots,m^{(L)}(n)\}.
\eeq

It is not difficult to see that when $L=1$, the policy reduces to the one described in Section~\ref{sec:DGFi}. 

Next, we establish the asymptotic optimality of the DGFi policy for $L>1$ and $K=1$. Let $\mathcal D$ denote a subset of $L$ cells and $\pi_{\mathcal D}$ the prior probability of hypothesis $H_{\mathcal D}$ (i.e, the target cells are given by $\mathcal{D}$). Define
\beq
\bea{l}
\label{eq:IstarL}
\displaystyle I_{\mathcal D}\triangleq \max\{\bar F_{\mathcal D},\bar G_{\mathcal D}\}, \\
\displaystyle I^*_L\triangleq \frac 1{\sum_{\mathcal D} \frac {\pi_{\mathcal D}}{I_{\mathcal D}}},
\ena
\eeq
where $I^*_L$ is again the optimal rate function of the Bayes risk as shown in the theorem below, and reduces to the one defined in \eqref{eq:Imstar} when $L=1$. 
\vspace{0.2cm}

\begin{theorem}
\label{th:optimality_policyL}
Let $R_L(\Gamma_{\footnotesize\mbox{DGFi}})$ and $R_L(\Gamma)$ be the Bayes risks under the DGFi policy and an arbitrary policy $\Gamma$, respectively.  For $K=1$, we have,
\beq
\label{eq:asymptotic_performanceL}
R_L(\Gamma_{\footnotesize\mbox{DGFi}}) \;\sim\; \frac{-c\log c}{I^*_L}\;\sim\;\inf_{\Gamma}\;{R(\Gamma)} \;.
\vspace{0.2cm}
\eeq
\end{theorem}

\begin{IEEEproof}
See Appendix D.
\end{IEEEproof}
\vspace{0.2cm}

For $K>1$, the stopping rule and the decision rule remain the same. For the selection rule, define
\beq
\label{FD}
\bea{l}
\displaystyle F_{\mathcal D}(\kappa)\triangleq \min\{{\kappa}\bar F_{\mathcal D},\;\min_{j\notin \mathcal D} D(f_j||g_j)\}.
\ena
\eeq

Similar to $F_m(\kappa)$ defined in (\ref{eq:fmk}), $F_{\mathcal D}(\kappa)$ can be viewed as the asymptotic increasing rate of $\Delta S_L(n)$ when the $L$ targets are given by set $\mathcal D$ and we probe those $\kappa$ cells with the $(L+1)^{th}$ to the $(L+\kappa)^{th}$ largest sum LLR. Similarly,
\beq
\label{GD}
\bea{l}
\displaystyle  G_{\mathcal D}(\kappa) \triangleq \min\{{\kappa}\bar G_{\mathcal D},\;\min_{j\in \mathcal D} D(g_j||f_j)\},
\ena
\eeq
which can be viewed as the asymptotic increasing rate of $\Delta S_L(n)$ when we probe the cells with the $(L-\kappa+1)^{th}$ to the $L^{th}$ largest sum LLR.

Let
\beq
\label{eq:kstar}
\bea{l}
\displaystyle k_{\mathcal D}^*\triangleq \arg\max_{k=0,1,\ldots,K} F_{\mathcal D}(K-k)+G_{\mathcal D}(k),
\ena
\eeq
which can be interpreted as the optimal number of target cells that should be probed at each time for maximizing the asymptotic increasing rate of $\Delta S_L(n)$. The selection rule of DGFi is thus given by
\beq
\phi(n)=\{m^{(L-k_{\mathcal D(n)}^*+1)}(n),\ldots, m^{(L-k_{\mathcal D(n)}^*+K)}(n)\},
\eeq
where
\beq
\mathcal D(n)=\{m^{(1)}(n),m^{(2)}(n),\ldots,m^{(L)}(n)\}.
\eeq

The asymptotic optimality of DGFi for $L>1$ and $K>1$ remains open. Following the same insight in the single-target case, however, we have strong belief of the following conjecture. 

\vspace{0.2cm}
\textsl{Conjecture 1:} The DGFi policy preserves its asymptotic optimality if
\beq
u_{\mathcal D}^*  \triangleq \arg\max_{u\in[0,K]}  F_{\mathcal D}(K-u)+G_{\mathcal D}(u)
\eeq
is an integer for all $\mathcal D$, where we allow the domain of $F_{\mathcal D}(\cdot)$ and $G_{\mathcal D}(\cdot)$ to be real numbers.  
\vspace{0.2cm}

\section{Comparison with the Chernoff Test}
\label{sec:compareChernoff}

In this section, we compare the performance of the proposed DGFi policy and the Chernoff test in terms of both computational complexity and sample complexity. 

\subsection{The Chernoff Test}
\label{sssec:Chernoff}

The Chernoff test has a randomized selection rule. Specifically, let $\mathbf q$ be a probability mass function over a set of $\omega$ available experiments $\left\{u_i\right\}_{i=1}^{\omega}$ that the decision maker can choose from. Note that in our case, ${\omega}=\binom{M}{K}$. For each hypothesis $m=1,2,\ldots,M$, the optimal action distribution is given by
\beq
\label{eq:selection_Chernoff}
\displaystyle \mathbf q^*_m=\arg\;\max_{\mathbf q}\;\min_{j\neq m}
\sum_{u_i}q^{u_i} D(p_{m}^{u_i}||p_j^{u_i})\;,
\eeq
where $p_j^{u_i}$ is the observation distribution under hypothesis $j$ when action $u_i$ is taken, and $\mathbf {q}^{u_i}$ is the $i$th element of $\mathbf q$ (i.e., the probability of choosing experiment $u_i$ under $\mathbf q$). The rationale behind \eqref{eq:selection_Chernoff} is a zero-sum game formulation of the problem, and the optimal mixed strategy $\mathbf{q}^*_m$ leads to a random observation that best differentiates $H_m$ from its closest alternative. 

The action at time~$n$ under the Chernoff test is drawn from a distribution $\mathbf q^*_{\hat i(n)}$, where $\hat{i}(n)$ is the ML estimate of the true hypothesis at time~$n$ based on past actions and observations. The stopping rule and the decision rule are the same as in (\ref{eq:stopping_policy1}), (\ref{eq:decision_policy1}).

The rate function of the Chernoff test $\Gamma_{\footnotesize\mbox{C}}$ under hypothesis $H_m$ is given by
\beq
\label{eq:ICh}
I_m(\Gamma_{\footnotesize\mbox{C}})=\;\min_{j\neq m}\sum_{u_i} {q^{*u_i}_{m}} D(p_{m}^{u_i}||p_j^{u_i}),
\eeq
which is the increasing rate of $\Delta S(n)$ under hypothesis $H_m$ when the Chernoff test is employed. The rate function of the Chernoff test under a given prior $\{\pi_m\}_{m=1}^M$ can be similary obtained as in~\eqref{eq:I}. 

We point out that in~\cite{chernoff1959sequential}, while proving $I_m(\Gamma_{\mbox{\footnotesize C}})$ equals the optimal rate $I^*_m$, Chernoff did not provide an explicit expression for $I^*_m$ or $I_m(\Gamma_{\mbox{\footnotesize C}})$.  Both were given, as in~\eqref{eq:ICh}, inexplicitly in terms of the optimizer $\mathbf q^*_m$ of the maximin problem in~\eqref{eq:selection_Chernoff}. Even for the problem studied here, a special case of that considered by Chernoff\footnote{Note that the asymptotic optimality of the Chernoff test requires the assumption of positive KL diverence between every pair of hypotheses under every experiment. This does not hold for the problem at hand. However, it can be shown that the Chernoff test preserves its asymptotic optimality in this case.}, solving for $\mathbf q^*_m$ numerically is computationally expensive (see a detailed analysis on computational complexity in the next subsection).  The explicit characterization of $I^*_m$ in~\eqref{eq:Imstar}, which equals to $I_m(\Gamma_{\mbox{\footnotesize DGFi}})$ in \eqref{eq:I_m} under the necessary and sufficient condition given in Theorem~\ref{th:optimality_policy3}, is a contribution of this work. 

\subsection{Comparison in Computational Complexity}
\label{sssec:computecomparison}
While both the Chernoff test and the DGFi policy are asymptotically optimal, i.e.,  $I (\Gamma_{\mbox{\footnotesize DGFi}})=I (\Gamma_{\mbox{\footnotesize C}})=I^*$, they differ drastically in computational complexity. Specifically, the Chernoff test can be expensive to compute especially when the number of hypotheses or the number of experiments is large. Consider the case of a single target ($L=1$). Computing the selection rule of the Chernoff test given in~(\ref{eq:selection_Chernoff}) requires solving $M$ minimax problems, each corresponding to a particular value of the ML estimate $\hat{i}(n)\in\{1,\ldots, M\}$. One efficient way of solving minimax problems is through linear programming, which takes polynomial time with respect to the number of variables and constraints. For this problem, the number of variables is $\binom{M}{K}$, which can be exponential in $M$ in the worst case. Calculating the rate function given in~\eqref{eq:ICh} requires the optimal selection distribution $\mathbf{q}^*_m$ for all $m$, thus bears similar computational complexity. For multi-target detection, the number of hypotheses is $\binom M L$, further increasing the complexity. 

The only computation involved in the selection rule of DGFi is (\ref{eq:fmk}), which requires $M$ summations each with $M-1$ elements. As a result, the computational time is $O(M^2)$, which is independent of $K$. Similarly, the computational complexity for calculating the rate function $I (\Gamma_{\mbox{\footnotesize DGFi}})$ is $O(M^2)$ as well. 

\subsection{Comparison in Sample Complexity}
\label{sssec:comparison}

In this subsection, we compare the performance of DGFi with that of the Chernoff test in the finite regime (i.e., when the sample cost $c$ is bounded away from $0$). 

Consider a uniform prior and exponentially distributed observations: $f_m\sim\exp(\lambda_f^{(m)})$ and $g_m\sim\exp(\lambda_g^{(m)})$. The KL divergences can be easily computed as follows. 
\begin{center}
$\bea{l}
\displaystyle D(g_m||f_m)=\log(\lambda_g^{(m)})-\log(\lambda_f^{(m)})+\frac{\lambda_f^{(m)}}{\lambda_g^{(m)}}-1\;, \vspace{0.1cm} \\
\displaystyle D(f_m||g_m)=\log(\lambda_f^{(m)})-\log(\lambda_g^{(m)})+\frac{\lambda_g^{(m)}}{\lambda_f^{(m)}}-1 \;.
\ena$
\end{center}

\begin{figure}[h!]
\centering
\includegraphics[width=2.0in]{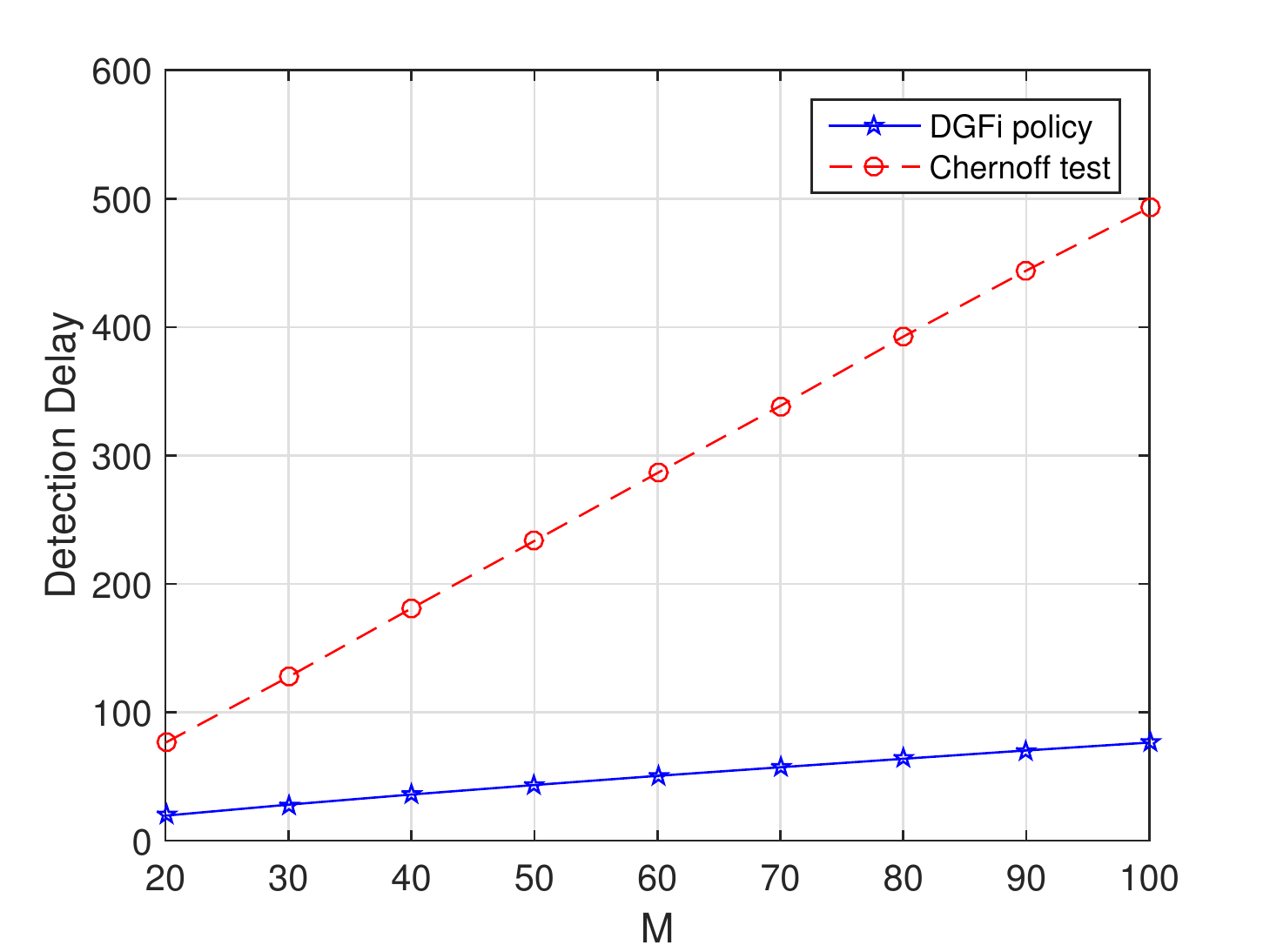} 
\includegraphics[width=2.0in]{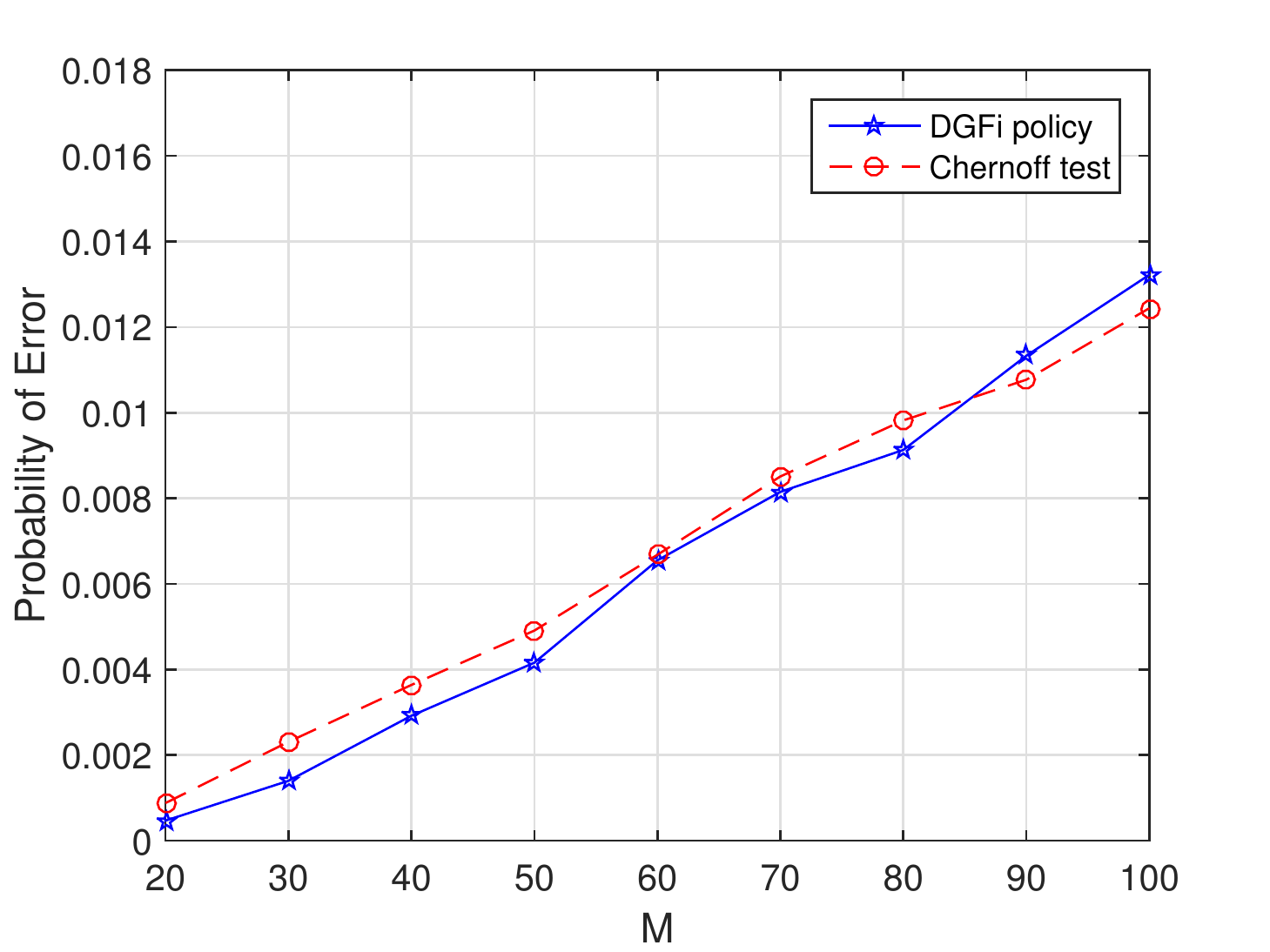} 
\includegraphics[width=2.0in]{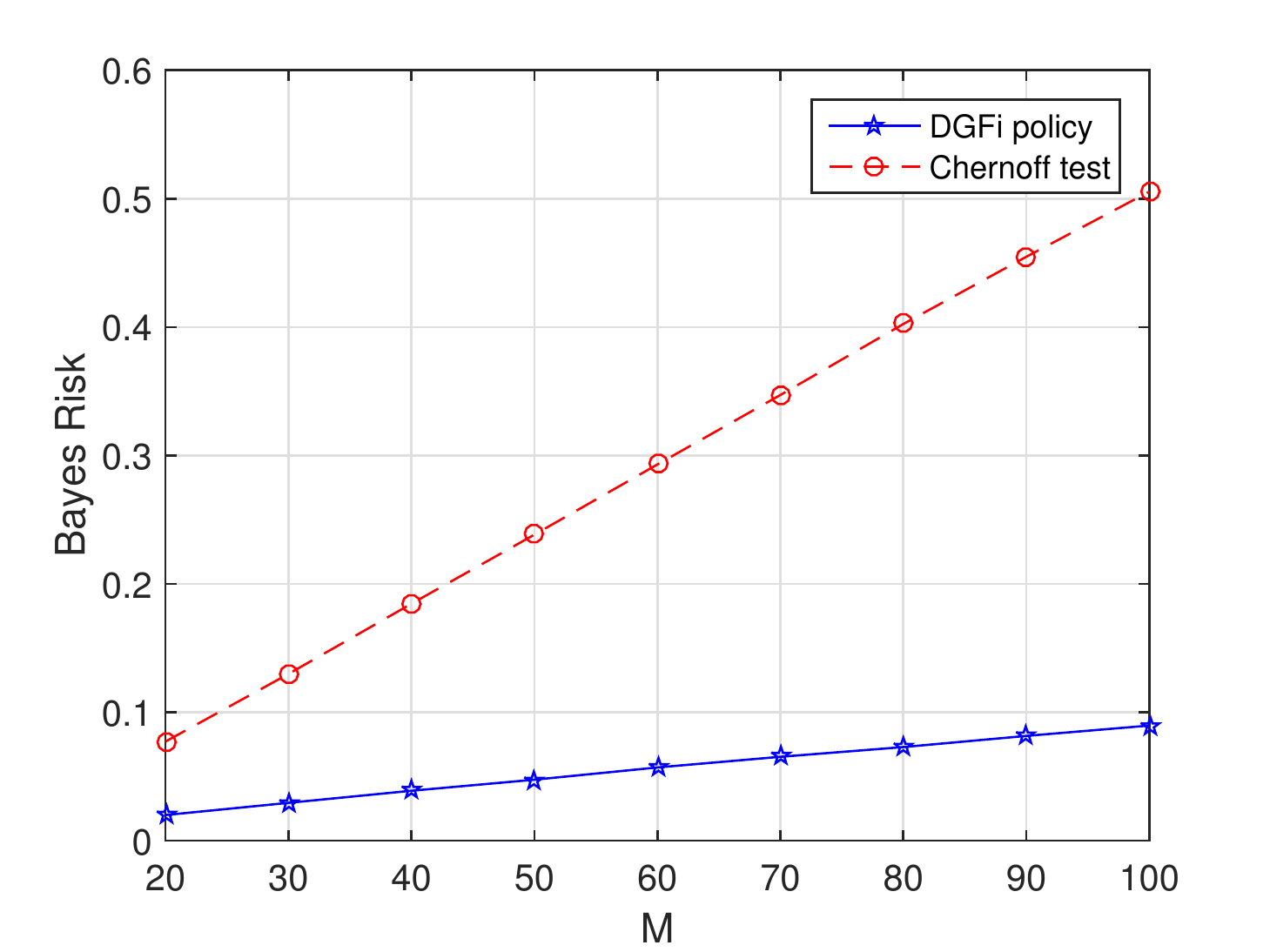}
\caption{Performance comparison ($K=1, \lambda_g^{(m)}=9+m,\lambda_f^{(m)}=0.0188,c=10^{-3}$).}
\label{fig_DGFi1}
\end{figure}

\begin{figure}[h!]
\centering
\includegraphics[width=2.0in]{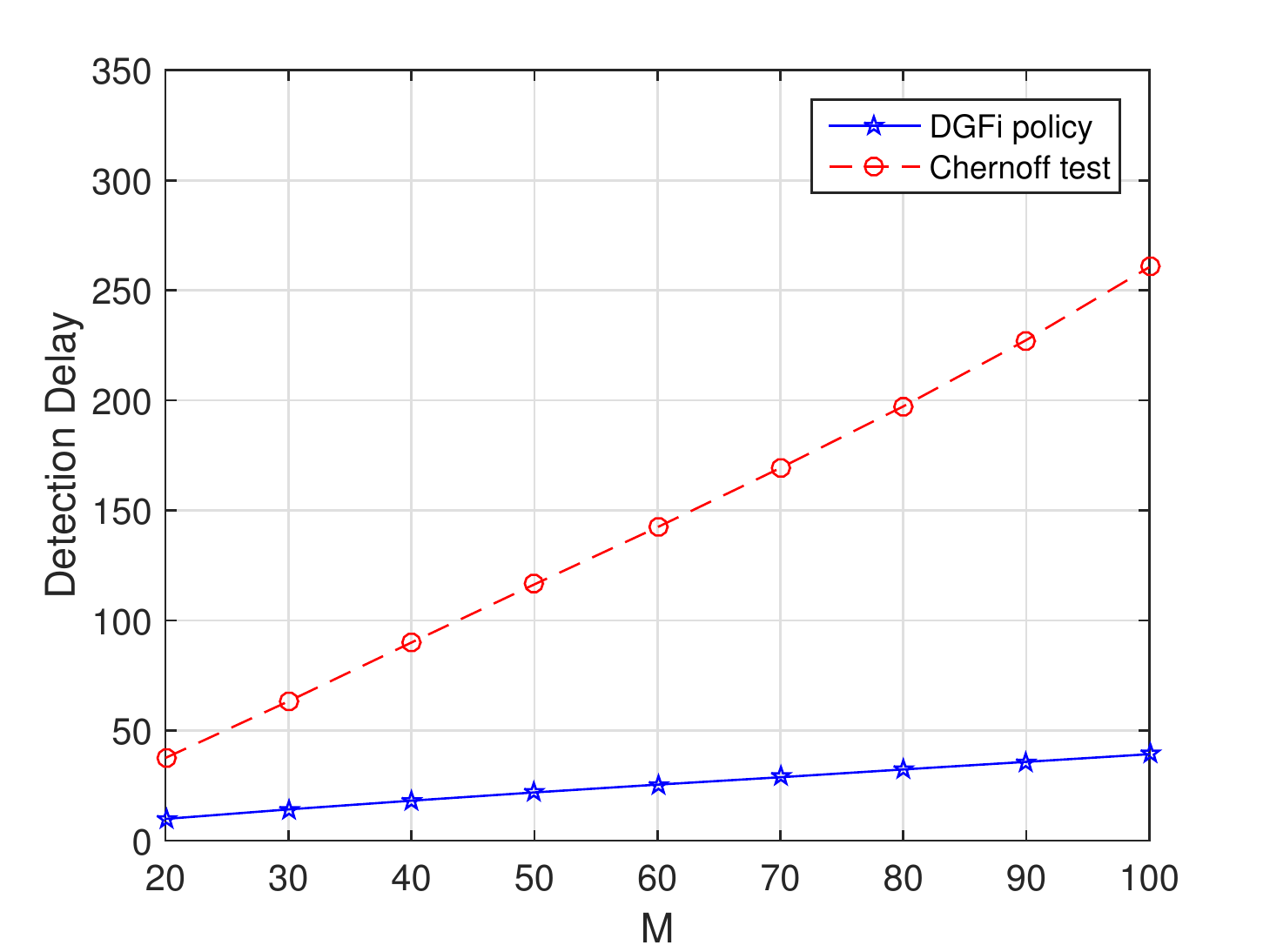} 
\includegraphics[width=2.0in]{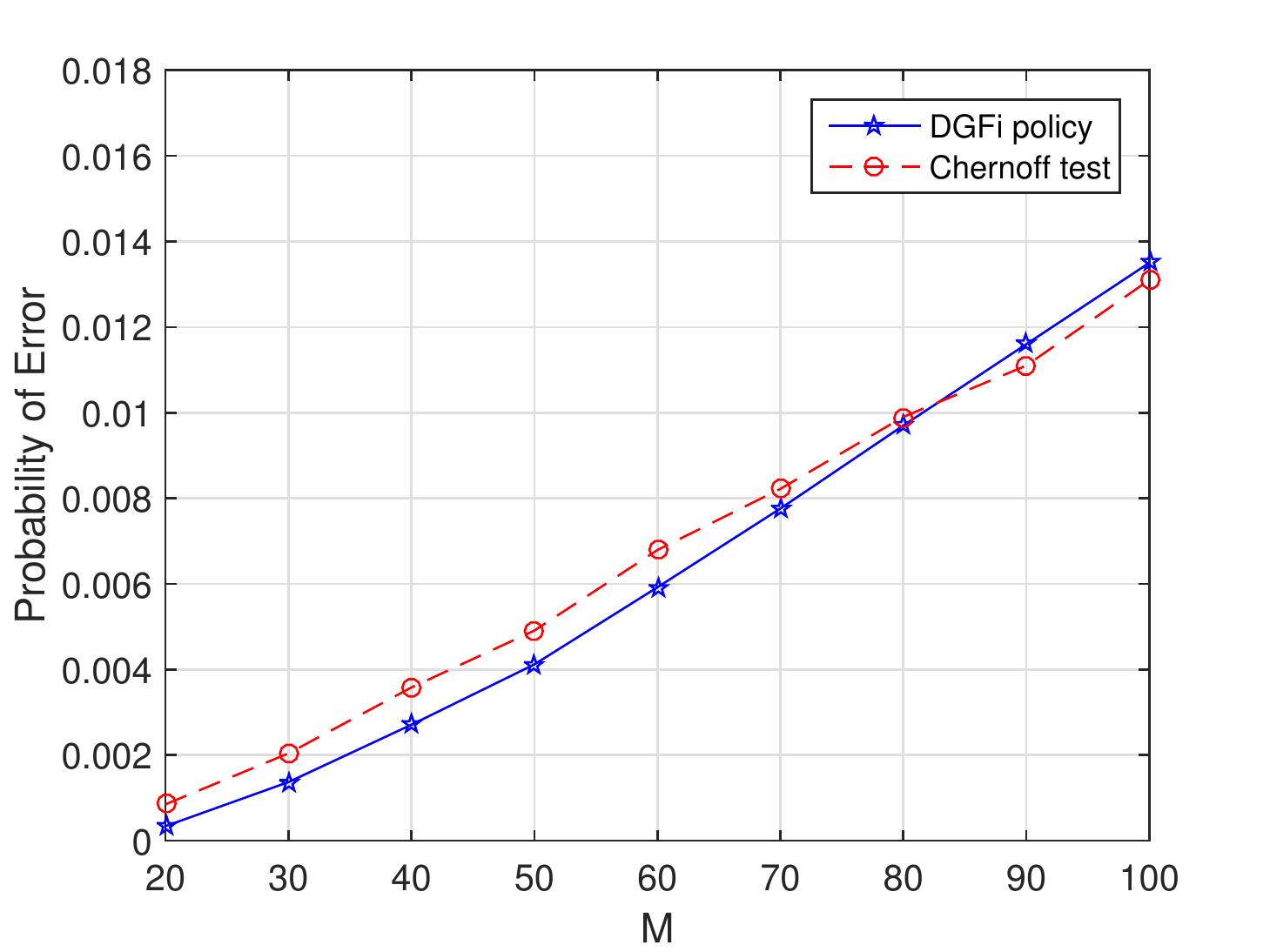} 
\includegraphics[width=2.0in]{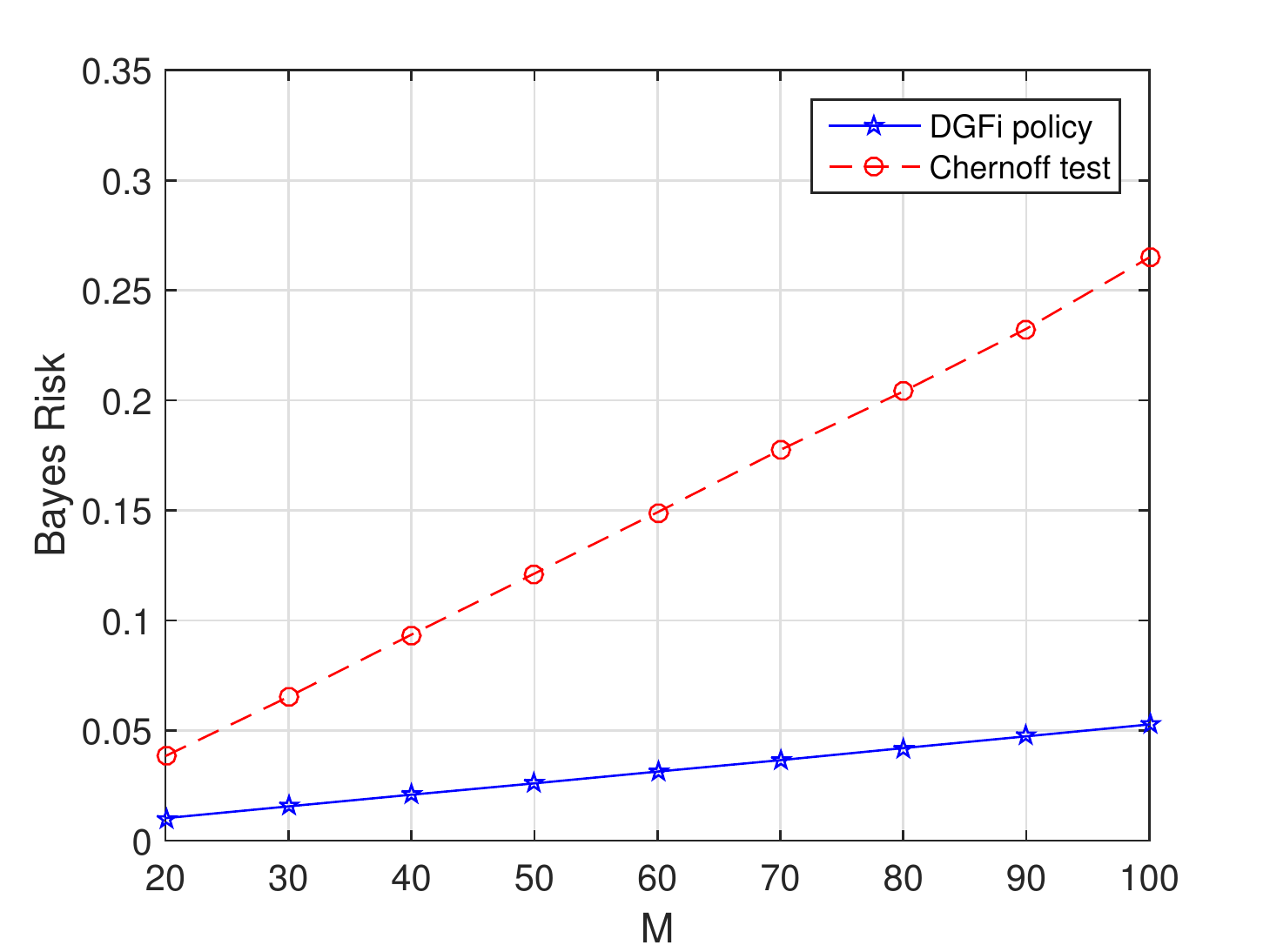}
\caption{Performance comparison ($K=2, \lambda_g^{(m)}=9+m,\lambda_f^{(m)}=0.0188,c=10^{-3}$).}
\label{fig_DGFi2}
\end{figure}

\begin{figure}[h!]
\centering
\includegraphics[width=2.0in]{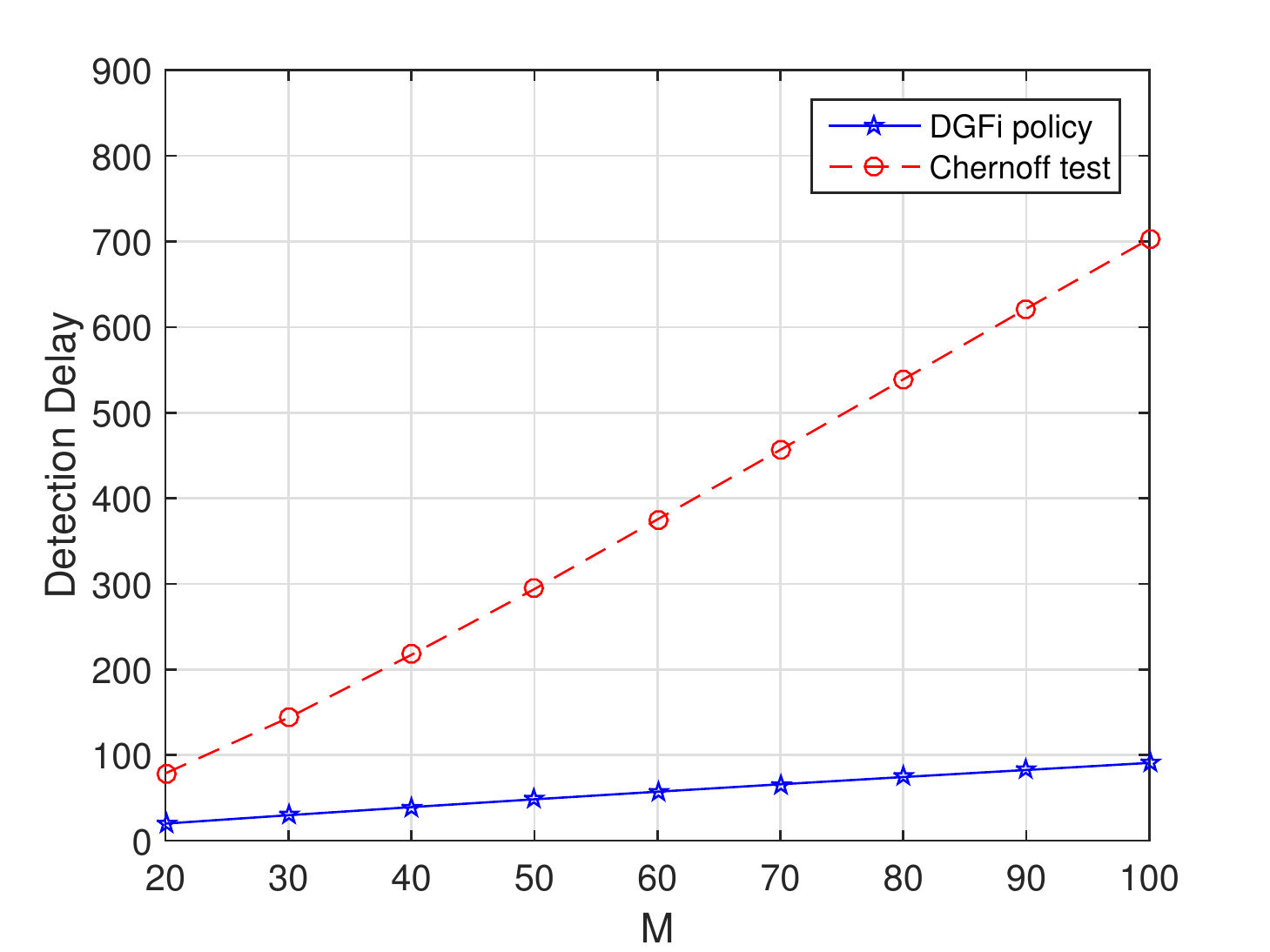} 
\includegraphics[width=2.0in]{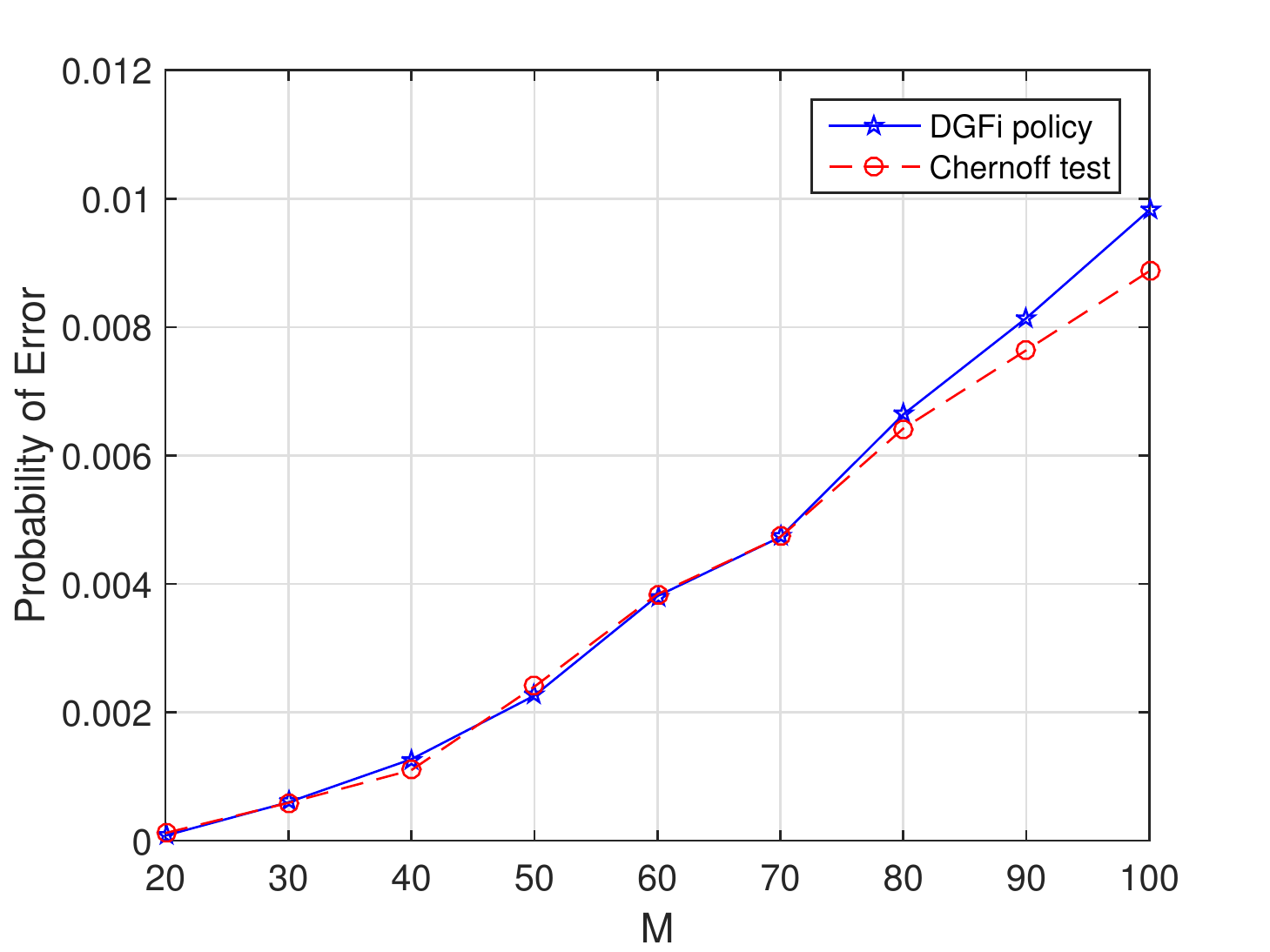} 
\includegraphics[width=2.0in]{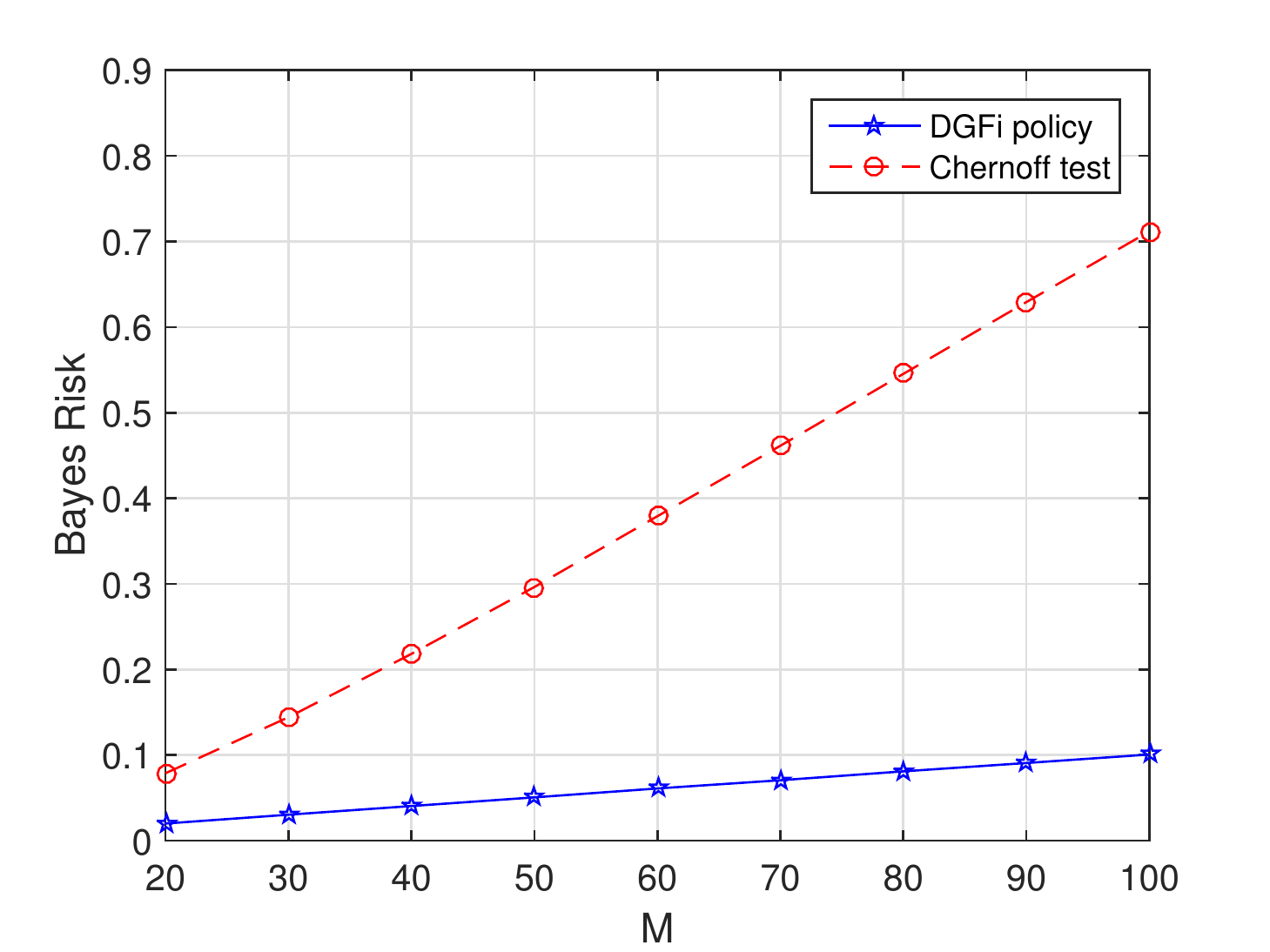}
\caption{Performance comparison ($L=2, K=1, \lambda_g^{(m)}=9+m,\lambda_f^{(m)}=0.0188,c=10^{-3}$).}
\label{fig_DGFiL}
\end{figure}

Shown in Fig.~\ref{fig_DGFi1} is the performance comparison between DGFi policy and Chernoff test for $L=1$ and $K=1$. The figure clearly demonstrates the significant reduction in detection delay and Bayes risk offered by the DGFi policy as compared with the Chernoff test. The performance gain increases drastically as $M$ increases.  The probability of errors for Chernoff test and DGFi policy are about the same order as shown. A similar comparison is observed in Fig.~\ref{fig_DGFi2} with $L=1,K=2$. The performance comparison for a case with multiple targets is shown in Fig.~\ref{fig_DGFiL} with $L=2, K=1$. 

Next, we provide an intuition argument for the superior finite-time performance of DGFi.  Consider a short horizon scenario where the sampling cost $c$ is sufficiently high such that $D(f||g)>-\log c$. This implies that each empty cell can be distinguished from the target with, on the average, a single probing to achieve the required accuracy as determined by $c$. We can cast this as the coupon collection problem, where each empty cell is a coupon  and the goal is to collect all $M-1$ coupons. Consider a special case where $K=1$ and all $f_i$ and $g_i$ are identical, i.e., $f_i\equiv f$ and $g_i\equiv g$. Assume that $D(f||g)>(M-1)D(g||f)$. In this case, the DGFi policy chooses, at each time, the cell with the second largest sum LLR whereas the Chernoff test randomly and uniformly chooses a cell from all but the one with the largest sum LLR at each time (this can be shown by solving \eqref{eq:selection_Chernoff}). Since Chernoff test  chooses empty cells with equal probability, based on results in coupon collectors problem, the expected probing time will be roughly $M\log M$. The DGFi policy, on the other hand, is deterministic and guaranteed to collect a new coupon at each time. The expected probing time is thus $M$.

\section{Conclusion}
\label{sec:conclusion}
The problem of detecting anomalies among a large number of heterogeneous processes was considered. A low-complexity deterministic test was developed and shown to be asymptotically optimal. Its finite-time performance and computational complexity were shown to be superior to the classic Chernoff test for active hypothesis testing, especially when the problem size is large. 

\section*{Appendix A: proof of Theorem~\ref{th:optimality_policy1}}
\label{thproof}

Throughout this section, we use the following notations. Let
\beq
N_j(n)\triangleq\sum_{t=1}^{n}{\mathbf{1}_j(t)}
\eeq
be the number of times that cell $j$ has been observed up to time~$n$. Let
\beq
\label{eq:Delta_S_m_j}
\Delta S_{m,j}(n)\triangleq S_m(n)-S_j(n) 
\eeq
be the difference between the observed sum of LLRs of cells $m$ and $j$. We also define
\beq
\label{eq:Delta_S_m}
\Delta S_m(n)\triangleq\min_{j\neq m} \Delta S_{m,j}(n) \;.
\eeq
As a result, we have:
\beq
\label{eq:Delta_Sn}
\Delta S(n)= S_{m^{(1)}(n)}(n)-S_{m^{(2)}(n)}(n)=\max_{m} \Delta S_m(n) \;.
\eeq

Without loss of generality we prove the theorem under hypothesis $H_m$. We define
\beq
\displaystyle \tilde{\ell}_k(i)=
\begin{cases} \ell_k(i)-D(g_k||f_k) \;,\;
                                                \mbox{if $k=m$,}   \vspace{0.3cm}\\
              \ell_k(i)+D(f_k||g_k) \;,\;
              \mbox{if $k\neq m$,}
\end{cases}
\eeq
which is a zero-mean random variable under hypothesis $H_m$. 

For the ease of presentation, we first provide the proof for the case of $K=1$. 

\subsection{Proof for $K=1$}
We first bound the error probability of DGFi as given below.
%
\begin{lemma}
\label{lemma:error_policy1}
If DGFi policy is used, then the error probability is upper bounded by:
\beq
\label{eq:Pe_bound_policy1}
P_e\leq (M-1)c \;.
\eeq
\end{lemma}
%
\begin{IEEEproof}
Let $\alpha_{m,j}=\mathbf{P}_m(\delta=j)$ for all $j\neq m$. Thus, $\alpha_m=\sum_{j\neq m}\alpha_{m,j}$.
By the definition of the stopping rule under DGFi (see~\eqref{eq:stopping_policy1}), accepting  $H_j$ is done when $\Delta S_j(n)\geq-\log c$ which implies $\Delta S_{j,m}\geq-\log c$.
Hence, for all $j\neq m$ we have:
\beq
\bea{l}
\alpha_{m,j}=\mathbf{P}_m\left(\delta=j\right)\vspace{0.2cm} \\
\leq\mathbf{P}_m\left(\Delta S_{j,m}(\tau)\geq-\log c\right)  \vspace{0.2cm} \\
\leq c\displaystyle\mathbf{P}_j\left(\Delta S_{j,m}(\tau)\geq -\log c\right)\leq c \;,
\ena
\eeq
where changing the measure in the second inequality follows by the fact that $\Delta S_{j, m}(\tau)\geq -\log c$. As a result,
\begin{center}
$\displaystyle\alpha_m=\sum_{j\neq m}\alpha_{m,j}\leq (M-1)c$\;
\end{center}
and \eqref{eq:Pe_bound_policy1} thus follows.
\vspace{0.3cm}
\end{IEEEproof}

Next we show that the expected detection time of DGFi is bounded by $-\log c/I_m (\Gamma_{\mbox{\footnotesize DGFi}})$ under hypothesis $H_m$. To show this, we partition the detection process into three stages, all defined by certain last passage times. The first stage is defined by the last passage time, denoted by $\tau_1$, that the maximum likelihood estimate is not the true hypothesis $H_m$.  The second stage defined by a last passage time $\tau_2$, indicates that the true hypothesis $H_m$ can be distinguished from at least one false hypothesis with sufficiently high accuracy. The third stage defined by last passage time $\tau_3$, indicates that $H_m$ can be distinguished from all the other $M-1$ hypotheses with sufficient accuracy.  The formal definitions of $\tau_1,\tau_2,\tau_3$ are give below:
\beq
\bea{l}
\label{eq:lastp}
\tau_1\triangleq\min \{t:\forall j\neq m,\forall n\ge t, S_m(n)\ge S_j(n)\} \vspace{0.2cm}\\ 
\tau_2\triangleq\min \{t:\exists j\neq m,\forall n\ge t, S_m(n) -S_j(n)\ge -\log c\} \vspace{0.2cm}\\
\tau_3\triangleq\min \{t:\forall j\neq m,\forall n\ge t, S_m(n) -S_j(n)\ge -\log c\}.
\ena
\eeq
Here, we assume that the selection rule of DGFi policy is implemented indefinitely, which means we probe the cells according to the selection rule of DGFi as given in~\eqref{eq:dgfk} indefinitely, while the stopping rule is disregarded. Note that $\tau_1,\tau_2,\tau_3$ are not stopping times since they depend on the future.  

 Since $\tau\le \tau_3$ based on the stopping rule of DGFi, it suffices to show $\tau_3$ is bounded by $-\log c/I_m (\Gamma_{\mbox{\footnotesize DGFi}})$ under hypothesis $H_m$. Let $n_2=\tau_2-\tau_1$ and $n_3=\tau_3-\tau_2$. In Lemma~\ref{lemma:tau_1_policy1} and Lemma~\ref{lemma:tau_3_policy1}, we show that $\tau_1$ and $n_3$ are sufficiently small with high probability.  In Lemma~\ref{lemma:tau_2_policy1} we show that the probability that $n_2$ is greater than $n$ decays exponentially with $n$ when $n$ is greater than  $-\log c/I_m (\Gamma_{\mbox{\footnotesize DGFi}})$.   Since $n_3=\tau_1+n_2+n_3$,  the expected detection time of DGFi is bounded by $-\log c/I_m (\Gamma_{\mbox{\footnotesize DGFi}})$ under hypothesis $H_m$ as desired.

\begin{lemma}
\label{lemma:S_j_S_m_N_j}
There exist constants $C>0$ and $\gamma>0$ such that for any fixed $0<q<1$, under any arbitrary policy, the following statements hold:
\beq
\label{eq:lemma:S_j_S_m_N_j}
\displaystyle\mathbf{P}_m\left(S_j(n)\geq S_m(n), N_j(n)\geq qn\right)\leq C e^{-\gamma n} \;,
\eeq
and
\beq
\label{eq:lemma:S_j_S_m_N_m}
\displaystyle\mathbf{P}_m\left(S_j(n)\geq S_m(n), N_m(n)\geq qn\right)\leq C e^{-\gamma n} \;,
\eeq
for $m=1, 2, \ldots, M$ and $j\neq m$.
\end{lemma}
\begin{IEEEproof}
We start with proving (\ref{eq:lemma:S_j_S_m_N_j}).
Note that $N_j(n), N_m(n)$ can take integer values $N_j(n)=\lceil qn\rceil, \lceil qn\rceil+1, ...n$, and $N_m(n)=0, ..., n$. Using the i.i.d. property of the observations across time yield:
\beq
\bea{l}
\displaystyle\mathbf{P}_m\left(S_j(n)\geq S_m(n), N_j(n)\geq qn\right)\vspace{0.2cm}\\
\leq\displaystyle\sum_{r=\lceil qn\rceil}^{n}\;\sum_{k=0}^{n}\mathbf{P}_m\left(\sum_{i=1}^{r}\ell_{j}(i)
                                                              +\sum_{i=1}^{k}-\ell_{m}(i)\geq 0\right)
\vspace{0.2cm}\\
\leq\displaystyle\sum_{r=\lceil qn\rceil}^{n}\;\sum_{k=0}^{n}\left[\mathbf{E}_m\left(e^{s\ell_j(1)}\right)\right]^{r}
    \left[\mathbf{E}_m\left(e^{s(-\ell_m(1))}\right)\right]^{k}
\ena
\eeq
where we have used the following generic Chernoff bound for a random variable $X$:
\beq
\label{gench}
\mathbf P(X\ge a) \le \frac{\E[e^{\lambda X}]}{e^{\lambda a}},
\eeq
where it is assume that the moment generating function $\E[e^{\lambda X}]$ exists locally in an interval around $\lambda=0$. Since the moment generating function is equal to one at $s=0$ and $\mathbf{E}_m(\ell_j(1))=-D(f_j||g_j)<0$, $\mathbf{E}_m(-\ell_m(1))=-D(g_m||f_m)<0$ are strictly negative, differentiating the MGFs of $\ell_j(1), \ell_m(1)$ with respect to $s$ yields strictly negative derivatives at $s=0$. As a result, there exist $s>0$ and $\gamma_1>0$ such that $\mathbf{E}_m\left(e^{s\ell_j(1)}\right)$, $\mathbf{E}_m\left(e^{s(-\ell_m(1))}\right)$ are strictly less than $e^{-\gamma_1}<1$. Hence, there exist $C>0$ and $\gamma=\gamma_1q>0$ such that
\beq
\bea{l}
\displaystyle\mathbf{P}_m\left(S_j(n)-S_m(n)\geq 0, N_j(n)\geq qn\right) \vspace{0.2cm}\\
\leq\displaystyle\sum_{r=\lceil qn\rceil}^{n} e^{-\gamma_1 r}  \sum_{k=0}^{n} e^{-\gamma_1 k}
\leq\displaystyle C e^{-\gamma n} \;.
\ena
\eeq
Note that \eqref{eq:lemma:S_j_S_m_N_m} can be proved with minor modifications. 
\vspace{0.1cm}
\end{IEEEproof}
\vspace{0.2cm}
\begin{lemma}
\label{lemma:tau_1_policy1}
If the selection rule of DGFi is implemented indefinitely, there exist $C>0$ and $\gamma>0$ such that
\beq
\label{eq:lemma_Pr_tau_1}
\mathbf{P}_m\left(\tau_1>n\right)\leq Ce^{-\gamma n}  \;,
\eeq
for $m=1, 2, \ldots, M$.
\end{lemma}
%
\begin{IEEEproof}
We focus on proving for $M>2$. Proving for $M=2$ is straightforward. Note that the event $\tau_1>n$ implies that there exists a time instant $t$ with $t\geq n$ such that $S_j(t)>S_m(t)$ for some $j\neq m$. Hence,
\beq\label{eq:tau1}
\bea{l}
\displaystyle\mathbf{P}_m\left(\tau_1>n\right)\leq\mathbf{P}_m\left(\max_{j\neq m}\;\sup_{t\geq n}\;\left(S_j(t)-S_m(t)\right)\geq 0 \right) \vspace{0.2cm} \\ \hspace{2cm}
\leq\displaystyle\sum_{j\neq m}\;\sum_{t=n}^{\infty}\mathbf{P}_m\left(S_j(t)\geq S_m(t)\right)\;.
\ena
\eeq
Following~\eqref{eq:tau1}, it suffices to show that there exist $C>0$ and $\gamma>0$ such that $\mathbf{P}_m\left(S_j(n)\geq S_m(n)\right)\leq Ce^{-\gamma n}$. \\

We next establish the required exponential decay. Let
\beq\label{eq:step1def}
\bea{l}
\displaystyle k_m=\frac{\max_{j\neq m} D(f_j||g_j)}{\min_{j\neq m} D(f_j||g_j)},\\
\displaystyle \underline{j_m}=\arg\min_{j\neq m} D(f_j||g_j),\\
\displaystyle\rho_m=\frac{1}{8(k_m+1)(M-2)}.
\ena
\eeq
Note that $0<\rho_m\leq 1/16$. Thus, we can write
\beq
\bea{l}
\label{eq:l_tau_1_policy1_Sj_geq_Sm}
\mathbf{P}_m\left(S_j(n)\geq S_m(n)\right) \vspace{0.2cm} \\ 
\leq\mathbf{P}_m\left(S_j(n)\geq S_m(n), N_j(n)<\rho_m n , N_m(n)<\rho_m n\right)
\vspace{0.2cm} \\ 
+\mathbf{P}_m\left(S_j(n)\geq S_m(n), N_j(n)\geq \rho_m n\right) \vspace{0.2cm} \\ 
+\mathbf{P}_m\left(S_j(n)\geq S_m(n), N_m(n)\geq \rho_m n\right).
\ena
\eeq
The second and the third terms on the RHS of~\eqref{eq:l_tau_1_policy1_Sj_geq_Sm} decay exponentially with $n$ by Lemma~\ref{lemma:S_j_S_m_N_j}. Thus, it remains to show that the first term decays exponentially with $n$ as well. Note that the event $(N_j(n)<\rho_m n, N_m(n)<\rho_m n)$ implies that at least $\tilde{n}=n-N_j(n)-N_m(n)\geq n\left(1-2\rho_m\right)$ times cells $j,m$ are not probed. We define $\widetilde{N}_r(n)$ as the number of times in which cell $r\neq j, m$ has been probed and cells $j,m$ have not been probed by time $n$. There exists a cell $r\neq j,m$ such that $\widetilde{N}_r(n)\geq\frac{\tilde{n}}{M-2}=\frac{n(1-2\rho_m)}{M-2}$. Hence, we can upper bound (\ref{eq:l_tau_1_policy1_Sj_geq_Sm}) as follows:
\beq\label{eq:Pm_Sj_Sm}
\bea{l}
\mathbf{P}_m\left(S_j(n)\geq S_m(n)\right) \vspace{0.2cm} \\ \hspace{0.3cm}

\displaystyle\leq\sum_{r\neq j,m}\mathbf{P}_m\bigg(\tilde{N}_r(n)>\frac{n(1-2\rho_m)}{M-2}, \vspace{0.2cm} \\ \hspace{0.3cm} N_j(n)<\rho_m n,  N_m(n)<\rho_m n\bigg)+2D e^{-\gamma_1 n},
\ena
\eeq
where the second and third terms on the RHS of~\eqref{eq:l_tau_1_policy1_Sj_geq_Sm} are upper bounded by $De^{-\gamma_1 n}$ (there exist such $D>0, \gamma_1>0$ by Lemma~\ref{lemma:S_j_S_m_N_j}), and the first term on the RHS of~\eqref{eq:l_tau_1_policy1_Sj_geq_Sm} is upper bounded by the first term (i.e., the summation term) on the RHS of~\eqref{eq:Pm_Sj_Sm}. Next, we show that each term in the summation decays exponentially with $n$ to get the desired result. 

Let $\tilde{t}^r_1, \tilde{t}^r_2, \ldots, \tilde{t}^r_{\tilde{N}_r(n)}$ be the indices for the time instants in which cell $r\neq j, m$ has been probed and cells $j, m$ have not been probed by time $n$. Let
\beq
\label{eq:zeta}
\displaystyle\zeta\triangleq\frac{1-2\rho_m}{2(M-2)} .
\eeq
Note that the event $S_j(\tilde{t}^r_{\zeta n})\leq S_r(\tilde{t}^r_{\zeta n})$ or $S_m(\tilde{t}^r_{\zeta n})\leq S_r(\tilde{t}^r_{\zeta n})$ must occur (otherwise, cell $j$ or $m$ will be probed). Hence\footnote{For the ease of presentation, throughout the proof we assume that $\zeta n$, $\rho_m n$ are integers. This assumption does not affect the exponential decay but only the exact value of $C>0$ in (\ref{eq:lemma_Pr_tau_1}) (since $\alpha n-1\leq\lfloor\alpha n\rfloor\leq\lceil\alpha n\rceil\leq\alpha n+1$ holds for all $\alpha\geq 0$ for all $n=0, 1, \ldots $).},
\beq\label{eq:tilde_Nr}
\bea{l}
\mathbf{P}_m\left(\tilde{N}_r(n)>\frac{n(1-2\rho_m)}{M-2},  \right. \vspace{0.2cm} \\ \hspace{1cm} 
\displaystyle\left.
N_j(n)<\rho_m n, N_m(n)<\rho_m n\right) \vspace{0.2cm} \\
=\displaystyle\sum_{q=0}^{n-\zeta n}\;\sum_{n'=0}^{\rho_m n}
\mathbf{P}_m\left(\sum_{i=1}^{n'}\ell_j(i)\leq \sum_{i=1}^{\zeta n+q}\ell_r(i)\right)\vspace{0.2cm} \\ \hspace{0.3cm}
+\displaystyle\sum_{q=0}^{n-\zeta n}\;\sum_{n'=0}^{\rho_m n}
\mathbf{P}_m\left(\sum_{i=1}^{n'}\ell_m(i)\leq \sum_{i=1}^{\zeta n+q}\ell_r(i)\right). \vspace{0.2cm} \\
\ena
\eeq
For upper bounding the first term on the RHS of~\eqref{eq:tilde_Nr} we write the sum LLRs as follows:

\beq
\bea{l}
 \displaystyle \sum_{i=1}^{\zeta n+q}\ell_r(i)+\sum_{i=1}^{n'}-\ell_j(i) \vspace{0.2cm} \\ \hspace{0.0cm}
\displaystyle  =\sum_{i=1}^{\zeta n+q}\tilde{\ell}_r(i)+\sum_{i=1}^{n'}\tilde{\ell}_j(i) \vspace{0.2cm} \\ \hspace{0.0cm}
\displaystyle-D(f_r||g_r)\left(\zeta n+q\right)+D(f_{n'}||g_{n'})n' \vspace{0.2cm} \\ 
\leq \displaystyle  \sum_{i=1}^{\zeta n+q}\tilde{\ell}_r(i)+\sum_{i=1}^{n'}-\tilde{\ell}_j(i)-D(f_{\underline{j_m}}||g_{\underline{j_m}})\left(\zeta n+q-k_mn'\right)
                    ,
\ena
\eeq
and by the definitions of $\zeta, k_m, \rho_m$ in~\eqref{eq:step1def} and~\eqref{eq:zeta}, we have
\begin{center}
$\bea{l}
\displaystyle\zeta n+q-k_mn'\geq \zeta n+q-k_mn'-(k_m+1)\left(\rho_m n-n'\right)\vspace{0.1cm}\\
\displaystyle=n\left(\zeta-(k_m+1)\rho_m\right)+q+n'\geq\frac{1}{4(M-2)}n+q+n'\vspace{0.1cm}\\
\displaystyle\geq\frac{1}{4(M-2)}(n+q+n')\;,
\vspace{0.1cm}\\
\ena$
\end{center}
for all $n'\leq\rho_m n$. Therefore,
\beq
\bea{l}
\displaystyle\sum_{i=1}^{\zeta n+q}\ell_r(i)+\sum_{i=1}^{n'}-\ell_j(i)\geq 0
\ena
\eeq
implies
\beq
\bea{l}
\displaystyle\sum_{i=1}^{\zeta n+q}\tilde{\ell}_r(i)+\sum_{i=1}^{n'}-\tilde{\ell}_j(i)\geq C_1\left(n+q+n'\right),
\ena
\eeq
where 
\beq
C_1=\frac{D(f_{\underline{j_m}}||g_{\underline{j_m}})}{4(M-2)}>0.
\eeq

Then we have
\beq\label{eq:Chernoff}
\bea{l}
\displaystyle\mathbf{P}_m\left(\sum_{i=1}^{n'}\ell_j(i)\leq \sum_{i=1}^{\zeta n+q}\ell_r(i)\right) \vspace{0.2cm}\\
\displaystyle\leq\mathbf{P}_m\left(\sum_{i=1}^{\zeta n+q}\tilde{\ell}_r(i)+\sum_{i=1}^{n'}-\tilde{\ell}_j(i)\geq C_1\left(n+q+n'\right)\right) \vspace{0.2cm}\\
\leq\displaystyle\left[\E_m\left(e^{s\tilde{\ell}_r(1)}\right)\right]^{\zeta n+q}
    \left[\E_m\left(e^{s(-\tilde{\ell}_j(1))}\right)\right]^{n'} \vspace{0.2cm} \\ \hspace{0.3cm}
\displaystyle\times e^{-sC_1\left(n+q+n'\right)} \vspace{0.2cm}\\
=\displaystyle\left[\E_m\left(e^{s\left(\tilde{\ell}_r(1)-C_1\right)}\right)\right]^{\zeta n+q}\left[\E_m\left(e^{s\left(-\tilde{\ell}_j(1)-C_1\right)}\right)\right]^{n'} \vspace{0.2cm} \\ \hspace{0.3cm}
\displaystyle\times e^{-sC_1\left(n-\zeta n\right)}
\;.
\ena
\eeq
for all $s>0$.

Since $\E_m(\tilde{\ell}_r(1)-C_1)=-C_1<0$ and $\E_m(-\tilde{\ell}_j(1)-C_1)=-C_1<0$ are strictly negative, by applying a similar argument as at the end of the proof of Lemma \ref{lemma:S_j_S_m_N_j}, there exist $s>0$ and $\gamma_2>0$ such that $\E_m\left(e^{(s\tilde{\ell}_r(1)-C_1)}\right)$, $\E_m\left(e^{s(-\tilde{\ell}_j(1)-C_1)}\right)$ and $e^{-sC_1}$ are strictly less than $e^{-\gamma_2}<1$.
Hence,
\beq\label{eq:Chernoff2}
\bea{l}
\displaystyle\mathbf{P}_m\left(\sum_{i=1}^{n'}\ell_j(i)\leq \sum_{i=1}^{\zeta n+q}\ell_r(i)\right)
    \leq\displaystyle e^{-\gamma_2\left(n+q+n'\right)},
\ena
\eeq
and
\beq
\bea{l}
\displaystyle\sum_{q=0}^{n-\zeta n}\;\sum_{n'=0}^{\rho_m n}
\mathbf{P}_m\left(\sum_{i=1}^{n'}\ell_j(i)\leq \sum_{i=1}^{\zeta n+q}\ell_r(i)\right) \vspace{0.2cm} \\ \hspace{0.3cm}
\displaystyle\leq e^{-\gamma_2 n} \sum_{q=0}^{n-\zeta n} e^{-\gamma_2 q}
\sum_{n'=0}^{\rho_m n} e^{-\gamma_2 n'}\leq C_2 e^{-\gamma_2 n} \;,
\ena
\eeq
where $C_2=\left(1-e^{-\gamma_2}\right)^{-2}$.\\

A similar technique can be applied to upper bound the second term on the RHS of \eqref{eq:tilde_Nr}.
\end{IEEEproof}
\vspace{0.2cm}

\begin{lemma}
\label{lemma:tau_2_policy1}
If the selection rule of DGFi is implemented indefinitely, then for every fixed $\epsilon>0$ there exist $C>0$ and $\gamma>0$ such that
\beq
\mathbf{P}_m\left(n_2>n\right)\leq C e^{-\gamma n} \;\;\;\; \forall n>-(1+\epsilon)\log c/I_m (\Gamma_{\mbox{\footnotesize DGFi}})\;,
\eeq
for all $m=1, 2, \ldots, M$.
\vspace{0.2cm}
\end{lemma}
%
\begin{IEEEproof} First, we consider the case where $I_m (\Gamma_{\mbox{\footnotesize DGFi}})>D(g_m||f_m)$. Note that cell $m$ is not observed for all $n\ge \tau_1$ in this case. Define $N_j'(\tau_1+t)=\sum_{i=\tau_1+1}^{\tau_1+t} 1_j(i)$ and $j^*(\tau_1+t)=\arg\max_j N_j'(\tau_1+t)D(f_j||g_j)$. Thus,
\beq
\bea{l}
\displaystyle \mathbf{P}_m(n_2>n) \vspace{0.2cm} \\ \hspace{0.3cm}
\displaystyle\le \mathbf{P}_m\left(\sup_{t\ge n}\sum_{i=\tau_1+1}^{\tau_1+t}\ell_{j^*(\tau_1+t)}(i)1_{j^*(\tau_1+t)}(i)\ge\log c\right).
\ena
\eeq
Since $t$ is the total number of observation from $\tau_1$ to $\tau_1+t$, by the definition of $j^*(t)$ we have
\beq
\bea{l}
\displaystyle  t=\sum_{j\neq m} N_j'(\tau_1+t)= \sum_{j\neq m}\frac{N_{j}'(\tau_1+t)D(f_{j}||g_{j})}{D(f_j||g_j)}  \vspace{0.2cm} \\ \hspace{0.3cm}
\displaystyle\le \sum_{j\neq m} \frac{N_{j^*(\tau_1+t)}'(\tau_1+t)D(f_{j^*(\tau_1+t)}||g_{j^*(\tau_1+t)})}{D(f_j||g_j)}.
\ena
\eeq
Let $\epsilon_1=I_m (\Gamma_{\mbox{\footnotesize DGFi}})\epsilon/(1+\epsilon)$. Since $I_m (\Gamma_{\mbox{\footnotesize DGFi}})=\sum_{j\neq m}1/D(f_j||g_j)$, we have 
\beq
\epsilon_1=\frac{\epsilon }{(1+\epsilon)\sum_{j\neq m}1/D(f_j||g_j)}.
\eeq
Then,
\beq
\bea{l}
\displaystyle \sum_{i=\tau_1+1}^{\tau_1+t}\ell_{j^*(\tau_1+t)}(i)1_{j^*(\tau_1+t)}(i)-\log c \\
\displaystyle = \sum_{i=\tau_1+1}^{\tau_1+t}\tilde \ell_{j^*(\tau_1+t)}(i)1_{j^*(\tau_1+t)}(i) \vspace{0.2cm} \\ \hspace{0.3cm}
\displaystyle-N_{j^*(\tau_1+t)}'(\tau_1+t)D(f_{j^*(\tau_1+t)}||g_{j^*(\tau_1+t)})-\log c \\
\displaystyle \le \sum_{i=\tau_1+1}^{\tau_1+t}\tilde \ell_{j^*(\tau_1+t)}(i)1_{j^*(\tau_1+t)}(i) \vspace{0.2cm} \\ \hspace{0.3cm}
\displaystyle-\frac{t}{\sum_{j\neq m}1/D(f_j||g_j)}-\log c\vspace{0.2cm} \\
\displaystyle \le\sum_{i=\tau_1+1}^{\tau_1+t}\tilde \ell_{j^*(\tau_1+t)}(i)1_{j^*(\tau_1+t)}(i)-tI_m (\Gamma_{\mbox{\footnotesize DGFi}})\vspace{0.2cm} \\ \hspace{0.3cm}
\displaystyle+tI_m (\Gamma_{\mbox{\footnotesize DGFi}})/(1+\epsilon) \\
\displaystyle \le \sum_{i=\tau_1+1}^{\tau_1+t}\tilde \ell_{j^*(\tau_1+t)}(i)1_{j^*(\tau_1+t)}(i) - t\epsilon_1
\ena
\eeq
for all $t\ge n >-(1+\epsilon)\log c/I_m (\Gamma_{\mbox{\footnotesize DGFi}})$. By applying the generic Chernoff bound given in~\eqref{gench}, it can be shown that there exists $\gamma_1>0$ such that $\mathbf P_m(\sum_{\tau_1+1}^{\tau_1+t}-\tilde \ell_{j^*(\tau_1+t)}(i) \ge t\epsilon_1)< e^{-\gamma_1 t}$ for all $t\ge n>-(1+\epsilon)\log c/I_m (\Gamma_{\mbox{\footnotesize DGFi}})$. Hence, there exist $C_1>0$ and $\gamma_1>0$ such that $\mathbf P_m(n_2>n)\le C_1e^{-\gamma_1 n}$ for all $n>-(1+\epsilon)\log c/I_m (\Gamma_{\mbox{\footnotesize DGFi}})$. A similar argument applies for case where $I_m (\Gamma_{\mbox{\footnotesize DGFi}})\le D(g_m||f_m)$.
\vspace{0.2cm}
\end{IEEEproof}

To show that $n_3$ is sufficiently small, we define a random variable $\Psi(t)$ as the dynamic range between sum LLRs of empty cells:
\beq
\displaystyle \Psi(t)\triangleq\max_{j\neq m}S_j(t)-\min_{j\neq m}S_j(t) . 
\eeq
Note that the dynamic range at time~$\tau_2$ can be viewed as a measure of the amount of information remains to gather in order to distinguish $H_m$ from any other false hypothesis. Lemma~\ref{lemma:DR_tau1_policy1} below shows that the dynamic range at time~$\tau_2$ is sufficiently small.
%

%
\begin{lemma}
\label{lemma:DR_tau1_policy1}
If the selection rule of DGFi is implemented indefinitely.
Then, for every fixed $\epsilon_1>0, \epsilon_2>0$ there exist $C>0$ and $\gamma>0$ such that
\beq
\bea{l}
\displaystyle\hspace{0.0cm}\mathbf{P}_m\left(\Psi(\tau_2)>\epsilon_1 n\right)\leq C e^{-\gamma n}, \;\vspace{0.2cm} \\ \hspace{2cm}
\displaystyle\forall n>-(1+\epsilon_2)\log c/I_m (\Gamma_{\mbox{\footnotesize DGFi}})\;
\ena
\eeq
for all $m=1, 2, \ldots, M$.
\vspace{0.2cm}
\end{lemma}
%
\begin{IEEEproof}
Note that
\beq\label{eq:pl_DR_tau1_policy1_1}
\bea{l}
\displaystyle\mathbf{P}_m\left(\Psi(\tau_2)>\epsilon_1 n\right) \vspace{0.2cm} \\
\leq\displaystyle\mathbf{P}_m\left(\tau_2>n\right) 
+\displaystyle\mathbf{P}_m\left(\Psi(\tau_2)>\epsilon_1 n, \tau_2\leq n\right) \vspace{0.2cm} \\
\ena
\eeq
Since $\tau_2=\tau_1+n_2$, applying Lemmas~\ref{lemma:tau_1_policy1},~\ref{lemma:tau_2_policy1} implies that the first term on the RHS of (\ref{eq:pl_DR_tau1_policy1_1}) decreases exponentially with $n$ for all $n>-(1+\epsilon_2)\log c/I_m (\Gamma_{\mbox{\footnotesize DGFi}})$ for every fixed $\epsilon_2>0$. It remains to show that the second term on the RHS of (\ref{eq:pl_DR_tau1_policy1_1}) decreases exponentially with $n$.
Let $\bar{j}=\arg\;\max_{j\neq m}S_j(\tau_2), \underline{j}=\arg\;\min_{j\neq m}S_j(\tau_2)$. Let $t_0$ be the smallest integer such that $S_{\underline{j}}(t)\leq S_{\bar{j}}(t)$ for all $t_0<t\leq\tau_2$.  As a result, $\Psi(\tau_2)>\epsilon_1 n$ implies
\begin{center}
$\displaystyle\sum_{t=t_0}^{\tau_2}\ell_{\bar{j}}(t)\mathbf{1}_{\bar{j}}(t)-\sum_{t=t_0}^{\tau_2}\ell_{\underline{j}}(t)\mathbf{1}_{\underline{j}}(t)
                        >\epsilon_1 n$\;.
\end{center}
Note that the second term on the RHS of (\ref{eq:pl_DR_tau1_policy1_1}) can be rewritten as:
\beq\label{eq:pl_DR_tau1_policy1_2}
\bea{l}
\displaystyle\mathbf{P}_m\left(\Psi(\tau_2)>\epsilon_1 n, \tau_2\leq n\right) \vspace{0.2cm} \\
=\displaystyle\mathbf{P}_m\left(\Psi(\tau_2)>\epsilon_1 n, \tau_2\leq n, t_0\geq\tau_1\right) \vspace{0.2cm} \\
\displaystyle+\mathbf{P}_m\left(\Psi(\tau_2)>\epsilon_1 n, \tau_2\leq n, t_0<\tau_1\right)
\ena
\eeq

First, we upper bound the first term on the RHS of (\ref{eq:pl_DR_tau1_policy1_2}). Note that for all $\tau_1\leq t_0<t\leq\tau_2$, we have $\mathbf{1}_{\underline{j}}(t)=0$. Hence, 

\beq
\label{eq:pl_DR_tau1_policy1_2_half}
\bea{l}
\displaystyle\sum_{t=t_0}^{\tau_2}\ell_{\bar{j}}(t)\mathbf{1}_{\bar{j}}(t)-\sum_{t=t_0}^{\tau_2}\ell_{\underline{j}}(t)\mathbf{1}_{\underline{j}}(t)=\sum_{t=t_0}^{\tau_2}\ell_{\bar{j}}(t)\mathbf{1}_{\bar{j}}(t)
                      \vspace{0.2cm} \\ \hspace{0.3cm}
\displaystyle=\sum_{t=t_0}^{\tau_2}\tilde{\ell}_{\bar{j}}(t)\mathbf{1}_{\bar{j}}(t)-D(f_{\bar j}||g_{\bar j})
                       \leq\sum_{t=t_0}^{\tau_2}\tilde{\ell}_{\bar{j}}(t)\mathbf{1}_{\bar{j}}(t)
                       
%
\ena
\eeq
Then, applying the generic Chernoff bound given in~\eqref{gench} completes the proof for this case.

Next, we upper bound the second term on the RHS of (\ref{eq:pl_DR_tau1_policy1_2}).
Let
$\epsilon_3\triangleq\frac{\epsilon_1}{4\max_j D(f_i||g_i)}>0$.
Note that
\beq\label{eq:pl_DR_tau1_policy1_3}
\bea{l}
\displaystyle\mathbf{P}_m\left(\Psi(\tau_2)>\epsilon_1 n, \tau_2\leq n, t_0<\tau_1\right) \vspace{0.2cm} \\
\leq\displaystyle\mathbf{P}_m\left(\tau_1>\epsilon_3 n\right) \vspace{0.2cm} \\
\displaystyle+\mathbf{P}_m\left(\Psi(\tau_2)>\epsilon_1 n, \tau_2\leq n, t_0<\tau_1, \tau_1\leq\epsilon_3 n\right) \;.
\ena
\eeq
The first term on the RHS of (\ref{eq:pl_DR_tau1_policy1_3}) decreases exponentially with $n$ by Lemma~\ref{lemma:tau_1_policy1}. Thus, it remains to show that the second term on the RHS of (\ref{eq:pl_DR_tau1_policy1_3}) decreases exponentially with $n$.
Note that $\Psi(\tau_2)>\epsilon_1 n$ implies
$\bea{l}
\sum_{t=t_0}^{\tau_1}\ell_{\bar{j}}\mathbf{1}_{\bar{j}}(t)

+\sum_{t=\tau_1+1}^{\tau_2}\ell_{\bar{j}}\mathbf{1}_{\bar{j}}(t)
                          >\epsilon_1 n. 
\ena$ Therefore, the second term on the RHS of (\ref{eq:pl_DR_tau1_policy1_3}) can be rewritten as:
\beq\label{eq:pl_DR_tau1_policy1_4}
\bea{l}
\displaystyle\mathbf{P}_m\left(\Psi(\tau_2)>\epsilon_1 n, \tau_2\leq n, t_0<\tau_1, \tau_1\leq\epsilon_3 n\right) \vspace{0.2cm} \\
\displaystyle\leq\mathbf{P}_m\left(\sum_{t=t_0}^{\tau_1}\ell_{\bar{j}}(t)\mathbf{1}_{\bar{j}}(t)

>\frac{\epsilon_1 n}{2}, 
\tau_2\leq n, t_0<\tau_1, \tau_1\leq\epsilon_3 n\right) \vspace{0.2cm} \\
\displaystyle+\mathbf{P}_m\left(\sum_{t=\tau_1+1}^{\tau_2}\ell_{\bar{j}}(t)\mathbf{1}_{\bar{j}}(t)

>\frac{\epsilon_1 n}{2}, 
\tau_2\leq n, t_0<\tau_1, \tau_1\leq\epsilon_3 n\right)
\ena
\eeq
The second term on the RHS of (\ref{eq:pl_DR_tau1_policy1_4}) decreases exponentially with $n$ using a similar argument as in (\ref{eq:pl_DR_tau1_policy1_2_half}).
Next, it remains to show that the first term on the RHS of (\ref{eq:pl_DR_tau1_policy1_4}) decreases exponentially with $n$.
Note that
\beq
\bea{l}
\displaystyle\displaystyle\sum_{t=t_0}^{\tau_1}\ell_{\bar{j}}(t)\mathbf{1}_{\bar{j}}(t)-\sum_{t=t_0}^{\tau_1}\ell_{\underline{j}}(t)\mathbf{1}_{\underline{j}}(t)
                    \vspace{0.2cm} \\
\displaystyle\leq\sum_{t=t_0}^{\tau_1}\tilde{\ell}_{\bar{j}}(t)\mathbf{1}_{\bar{j}}(t)-\sum_{t=t_0}^{\tau_1}\tilde{\ell}_{\underline{j}}(t)\mathbf{1}_{\underline{j}}(t)
                       
                        +\max_jD(f_j||g_j)\tau_1
                    \vspace{0.2cm} \\
\displaystyle\leq\sum_{t=t_0}^{\tau_1}\left[\tilde{\ell}_{\bar{j}}(t)\mathbf{1}_{\bar{j}}(t)
                        -\tilde{\ell}_{\underline{j}}(t)\mathbf{1}_{\underline{j}}(t)\right]
                        +\frac{\epsilon_1}{4}n
                    \vspace{0.2cm} \\
\ena
\eeq
for all $\tau_1\leq\epsilon_3 n$.\\
As a result,
\beq
\bea{l}
\displaystyle\displaystyle\sum_{t=t_0}^{\tau_1}\ell_{\bar{j}}(t)\mathbf{1}_{\bar{j}}(t)
                        -\ell_{\underline{j}}(t)\mathbf{1}_{\underline{j}}(t) >\frac{\epsilon_1}{2} n
\ena
\eeq
implies
\beq
\bea{l}
\displaystyle\sum_{t=t_0}^{\tau_1}\left[\tilde{\ell}_{\bar{j}}(t)\mathbf{1}_{\bar{j}}(t)
                        -\tilde{\ell}_{\underline{j}}(t)\mathbf{1}_{\underline{j}}(t)\right]
                        >\frac{\epsilon_1}{4}n
                    \vspace{0.2cm} \\
\ena
\eeq
for all $\tau_1\leq\epsilon_3 n$. Applying the generic Chernoff bound given in~\eqref{gench}, we arrive at the lemma.

\end{IEEEproof}
\vspace{0.2cm}

\begin{lemma}
\label{lemma:tau_3_policy1}
If the selection rule of DGFi is implemented indefinitely, then for every fixed $\epsilon>0$ there exist $C>0$ and $\gamma>0$ such that
\beq\label{eq:l_tau3}
\mathbf{P}_m\left(n_3>n\right)\leq C e^{-\gamma n} \;\;\;\; \forall n>-\epsilon\log c/I_m (\Gamma_{\mbox{\footnotesize DGFi}})\;,
\eeq
for all $m=1, 2, \ldots, M$.
\vspace{0.2cm}
\end{lemma}
%
\begin{IEEEproof}
To prove the Lemma, we first define $\tau_3^j\triangleq\max\{t:\forall n\ge t, S_m(n)-S_j(n)\ge -\log c\}$ and $N_3^j$ as the total number of observations that the decision maker collected from cell $j$ between $\tau_2$ and $\tau_3^j$. Since $n_3\le \sum_j N_3^j$ and $\tau_3=\max_j \tau_3^j$, we only need to show that $\mathbf {P}_m(N_3^j>n)$ decays exponentially with $n$. We can write $\mathbf P_m(N_3^j>n)$ as follows:
\beq
\bea{l}
\displaystyle \mathbf P_m(N_3^j>n)\le \mathbf {P}_m\left(\Psi(\tau_2)>n\frac{\min_j D(f_j||g_j)}{2}\right) \vspace{0.2cm}\\ \hspace{0.5cm}
 \displaystyle+\mathbf {P}_m\left(N_3^j>n|\Psi(\tau_2)\le n\frac{\min_j D(f_j||g_j)}{2}\right)
\ena
\eeq
Lemma 6 provides the desired decay for the first term on the RHS. We next show the desired decay for the second term. Let $t_1,t_2,\ldots$ denote the time indices when cell $j$ is observed between $\tau_2$ and $\tau_3^j$. We can write:
\beq
\bea{l}
\displaystyle \mathbf {P}_m\left(N_3^j>n|\Psi(\tau_2)\le n\frac{\min_j D(f_j||g_j)}{2}\right) \\
\displaystyle\le \mathbf {P}_m\left(\inf_{r>n} \sum_{i=1}^r -\ell_j(t_i)<n\frac{\min_j D(f_j||g_j)}{2}\right) \\
\displaystyle\le \mathbf {P}_m\left( \sum_{i=1}^r \tilde \ell_j(t_i)> r\frac{\min_j D(f_j||g_j)}{2}\right).
\ena
\eeq
Using the i.i.d. property of $\tilde \ell_j(t_i)$ yields:
\beq
\mathbf {P}_m\left( \sum_{i=1}^n \tilde \ell_j(t_i)> n\frac{\min_j D(f_j||g_j)}{2}\right)<C_3 e^{-\gamma n}
\eeq
for some $C_3,\gamma_3$ which completes the proof.
\vspace{0.2cm}
\end{IEEEproof}

The following Lemma provides an upper bound on the detection time when DGFi policy is implemented.

\begin{lemma}
\label{lemma:expected_time_policy1}
If DGFi policy is implemented, then the expected detection time $\tau$ is upper bounded by:
\beq\label{eq:lemma_expected_time}
\E_m(\tau)\leq -\left(1+o(1)\right)\frac{\log(c)}{I_m (\Gamma_{\mbox{\footnotesize DGFi}})} \;,
\eeq
for $m=1, \ldots, M$.
\vspace{0.2cm}
\end{lemma}
%
\begin{IEEEproof}
Since the actual detection time under DGFi is upper bounded by: $\tau\leq\tau_3=\tau_1+n_2+n_3$, combining Lemmas~\ref{lemma:tau_1_policy1},~\ref{lemma:tau_2_policy1} and~\ref{lemma:tau_3_policy1} proves the statement.
\vspace{0.2cm}
\end{IEEEproof}

Combining Lemma \ref{lemma:error_policy1} and Lemma \ref{lemma:expected_time_policy1}, Theorem 1 follows for the case of $K=1$. 

\subsection{Proof for $K>1$}

We focus on the case where $F_m(K)>D(g_m||f_m)+F_m(K-1)$. The case where the inequality is reversed can be proven with minor modifications. 

We consider the balanced case and the unbalanced case separately. For the balanced case, the proof in Subsection A directly applies. For the unbalanced case, the proof has to be constructed differently. This is because in the unbalanced case, there is a process with a sufficiently small information acquisition rate $D(f_j||g_j)$ such that it becomes the bottleneck of the detection process and determines the asymptotic increasing rate of $\Delta S(n)$. Directly bounding the dynamic range of all sum LLR trajectories is no longer tractable. Instead, the proof is built upon the analysis of the trajectory of the sum LLR with the smallest expected increment. In particular, we recognize that the key in handling the imbalance in the information acquisition rates among empty cells is to define a last passage time as the last time at which the empty cell with the smallest $D(f_j||g_j)$ is not probed and then analyze, separately, the detection process before and after this last passage time.

The proof proceeds as follows. First, by directly applying Lemma 2, the error probability under DGFi is $O(c)$. Then, we show that the expected detection time of DGFi is bounded. Similar to the case of $K=1$, we partition the detection process into three stages with minor modifications. The first and the third stage are defined by the same last passage times $\tau_1$ and $\tau_3$ given in~\eqref{eq:lastp}. The second stage, however, is defined differently by $\tilde \tau_2$, indicates that the sum LLR of the cell with the smallest KL divergence is smaller than $-\log c$. By directly applying Lemmas 3 and 4, we show that $\tau_1$ is sufficiently small with high probability. 

Then, we prove the following Lemmas to show that $\tau_3$ is bounded by $\frac{-\log (c)}{I_m (\Gamma_{\mbox{\footnotesize DGFi}})}$. Lemma \ref{lemma:lower} states that the largest observed sum LLR among the empty cells is sufficiently large as required with high probability. Lemma~\ref{lemma:upper} states that the smallest observed sum LLR among the empty cells is sufficiently small as required with high probability. Lemma~\ref{lemma:dr} shows the difference between the $(K+1)^{th}$ largest sum LLR and the $M^{th}$ largest sum LLR is sufficiently small as required with high probability. Lemma \ref{lemma:slow} states that the sum LLR of the cell with the smallest KL divergence is sufficiently small (which will determine the rate function function for the search in this case) with high probability. Lemma \ref{lemma:other} shows that the sum LLR of other cells are smaller than that of the cell with the smallest KL divergence at time when $t>\tilde \tau_2$. Finally, Lemma \ref{lem:tau3} upper bounds the last passage time $\tau_3$.

Define
\beq
U(n)\triangleq\min_{j\neq m} S_j(n),\;L(n)\triangleq\max_{j\neq m} S_j(n),
\eeq

\beq
\Psi^{j_2}_{j_1}(n)\triangleq S_{m^{(j_2)}(n)}(n)-S_{m^{(j_1)}(n)}(n),
\eeq

\beq \underline j(t)\triangleq\arg\min_{j\neq m} N_j(t)D(f_j||g_j), \eeq

 \beq \bar j(t)\triangleq\arg\max_{j \neq m} N_j(t)D(f_j||g_j),\eeq
 
\beq j'=\arg \min_{j\neq m} D(f_j||g_j). \eeq

\begin{lemma}
\label{lemma:lower}
For any selection rule, $\forall t, \forall \epsilon>0$, there exist $C,\gamma>0$ such that
\beq
\mathbf {P}_m(L(t)<-tK \bar F_m-n\epsilon)<C e^{-\gamma n} \quad \forall n>t. \vspace{0.2cm}
\eeq
\end{lemma}
\begin{IEEEproof}
Note that
\beq
\mathbf {P}_m(L(t)<-tK \bar F_m-n\epsilon)\le \mathbf {P}(S_{\underline j(t)}(t)<-tK \bar F_m-n\epsilon),
\eeq
and
\beq
S_{\underline j(t)}(t)= -N_{\underline j(t)}(t)D(f_{\underline j(t)}||g_{\underline j(t)})+\sum_{i=1}^t \tilde l_{\underline j(t)}(i) 1_{\underline j(t)}(i).
\eeq
Since $Kt$ is the total number of observations by time $t$, by the definition of $\underline j(t)$ we have
\beq
\bea{l}
\displaystyle Kt=\sum_{j} N_j(t)= \sum_{j}\frac{N_{j}(t)D(f_{j}||g_{j})}{D(f_j||g_j)} \\ \hspace{0.5cm}
\displaystyle \ge \sum_j \frac{N_{\underline j(t)}(t)D(f_{\underline j(t)}||g_{\underline j(t)})}{D(f_j||g_j)}.
\ena
\eeq
Hence,
\beq
N_{\underline j(t)}(t)D(f_{\underline j(t)}||g_{\underline j(t)}) \le Kt \cdot \frac{1}{\sum_j 1/D(f_j||g_j)}=t\cdot K\bar F_m.
\eeq
Therefore,
\beq
S_{\underline j(t)}(t)<-tK\bar F_m-n\epsilon
\eeq
implies
\beq
\sum_{i=1}^t \tilde l_{\underline j(t)}(i) 1_{\underline j(t)}(i) < -n\epsilon.
\eeq
Then, applying the generic Chernoff bound completes the proof. \vspace{0.3cm}
\end{IEEEproof}

\begin{lemma}
\label{lemma:upper}
For any selection rule, $\forall t, \forall \epsilon$, there exist $C,\gamma>0$ such that
\beq
\mathbf {P}_m(U(t)>-t K \bar F_m+n\epsilon)<C e^{-\gamma n} \quad \forall n>t. \vspace{0.2cm}
\eeq
\end{lemma}

\begin{IEEEproof}
The proof follows similarly with Lemma~\ref{lemma:lower}.
\end{IEEEproof}

\vspace{0.3cm}

\begin{lemma}
\label{lemma:dr}
If DGFi policy is implemented, $\forall t, \forall \epsilon$, there exist $C,\gamma>0$ such that
\beq
\mathbf {P}_m(\Psi_{M}^{K+1}(t)>\max_{j\neq m}D(f_j||g_j)+n\epsilon)<C e^{-\gamma n} \quad \forall n>t. \vspace{0.2cm}
\eeq
\end{lemma}
\begin{IEEEproof}
We prove by induction with respect to $t$. When $t=1$, using the generic Chernoff bound completes the induction base. If the statement is true for $t-1$, then for $t$ we have
\begin{equation}
\label{eq:dr1}
\bea{l}
\displaystyle \mathbf {P}_m(\Psi_{M}^{K+1}(t)>\max_{j\neq m}D(f_j||g_j)+n\epsilon)\\
= \mathbf {P}_m(\Psi_{M}^{K+1}(t)>\max_{j\neq m}D(f_j||g_j)+n\epsilon, \vspace{0.2cm} \\ \hspace{2.0cm}
\displaystyle  m^{(M)}(t)=m^{(M)}(t-1))\\
+\mathbf {P}_m(\Psi_{M}^{K+1}(t)>\max_{j\neq m}D(f_j||g_j)+n\epsilon,\vspace{0.2cm} \\ \hspace{2.0cm}
\displaystyle m^{(M)}(t)\ne m^{(M)}(t-1)).
\ena
\end{equation}

For the first term on the RHS, we have
\beq
\label{eq:dr2}
\bea{l}
\displaystyle \mathbf {P}_m(\Psi_{M}^{K+1}(t)>\max_{j\neq m}D(f_j||g_j)+n\epsilon,\vspace{0.2cm} \\ \hspace{2.0cm}
\displaystyle m^{(M)}(t)=m^{(M)}(t-1))\\
\displaystyle\le \mathbf {P}_m(\Psi_{M}^{K+1}(t-1)>\max_{j\neq m}D(f_j||g_j)+\frac {n\epsilon}2,\vspace{0.2cm} \\ 
\displaystyle m^{(M)}(t)=m^{(M)}(t-1)\mbox{ or } l_{m^{(K+1)}(t-1)}(t)<-\frac{n\epsilon}{2},\vspace{0.2cm}\\ \hspace{2.0cm}
\displaystyle m^{(M)}(t)=m^{(M)}(t-1))\\
\displaystyle\le \mathbf {P}_m(\Psi_{M}^{K+1}(t-1)>\max_{j\neq m}D(f_j||g_j)+\frac {n\epsilon}2) \vspace{0.2cm}\\ \hspace{0.5cm}
\displaystyle+\mathbf {P}_m(l_{m^{(K+1)}(t-1)}(t)<-\frac{n\epsilon}{2})\vspace{0.2cm}\\ 
\displaystyle \le C_1 e^{-\gamma_1 n},
\ena
\eeq
where the first term can be bounded using assumptions on $t-1$ and the second term can be bounded using the generic Chernoff bound.

For the second term on the RHS of~\eqref{eq:dr1}, we have
\begin{equation}
\label{eq:dr3}
\bea{l}
\hspace{-0.2cm}\displaystyle\mathbf {P}_m(\Psi_{M}^{K+1}(t)>\max_{j\neq m}D(f_j||g_j)+n\epsilon,r^{(M)}(t)\ne r^{(M)}(t-1))\vspace{0.2cm}\\ \hspace{0.5cm}
\displaystyle
\le \mathbf {P}_m(l_{m^{(M)}(t)}(t)>\max_{j\neq m}D(f_j||g_j)+n\epsilon) \vspace{0.2cm} \\
\displaystyle\le \mathbf {P}_m(\tilde l_{m^{(M)}(t)}(t)>n\epsilon)< C_2 e^{-\gamma_2 n}.
\ena
\end{equation}
Combining~\eqref{eq:dr1},~\eqref{eq:dr2},~\eqref{eq:dr3} completes the proof.
\end{IEEEproof}

\begin{lemma}
\label{lemma:slow}
If DGFi policy is implemented, $\forall t>\tau_1, \forall \epsilon$, there exist $C,\gamma>0$ such that
\beq
\bea{l}
\hspace{-0.2cm} \mathbf {P}_m(S_{j'}(t)-S_{j'}(\tau_1)>-(t-\tau_1) D(f_{j'}||g_{j'})+n\epsilon)<C e^{-\gamma n} \vspace{0.2cm}\\ \hspace{3.5cm}
\displaystyle\forall n>t.
\ena
\eeq
\end{lemma}
\begin{IEEEproof}
Define $t_0$ as the smallest integer such that cell $j'$ is observed at time $i$ for all $t_0<i\le t$. Then, by our selection rule, cell $j'$ is the one of the top $K$ sum LLRs at time $t_0$. Then, by applying $t=t_0$ to Lemma~\ref{lemma:dr} we have
\beq
\mathbf {P}_m(U(t_0)-S_{j'}(t_0)<-n\epsilon)<C_1 e^{-\gamma_1 n} \quad \forall n>t_0,
\eeq
for some $C_1,\gamma_1$. Substituting $t=t_0$ in Lemma~\ref{lemma:upper} we have:
\beq
\mathbf {P}_m(U(t_0)>-t_0K \bar F_m+n\epsilon)<C_2 e^{-\gamma_2 n} \quad \forall n>t_0 ,
\eeq
for some $C_2,\gamma_2$. Hence,
\beq
\mathbf {P}_m(S_{j'}(t_0)>-t_0K \bar F_m+n\epsilon)<C_3 e^{-\gamma_3 n} \quad \forall n>t_0 ,
\eeq
for some $C_3,\gamma_3$. Then, by the definition of $t_0$ and using the generic Chernoff bound we have
\beq
\bea{l}
\mathbf {P}_m(S_{j'}(t)-S_{j'}(t_0)>-(t-t_0)D(f_{j'}||g_{j'})+n\epsilon)\vspace{0.2cm}\\ \hspace{0.5cm}
\displaystyle <C_4 e^{-\gamma_4 n} \hspace{1cm} \forall n>(t-t_0).
\ena
\eeq
Since $K\bar F_m>D(f_{j'}||g_{j'})$, we have:
\beq
\bea{l}
\mathbf {P}_m(S_{j'}(t)-S_{j'}(\tau_1)>-(t-\tau_1) D(f_{j'}||g_{j'})+n\epsilon)\vspace{0.2cm}\\ \hspace{0.5cm}<C_5 e^{-\gamma_5 n} \hspace{1cm} \forall n>t
\ena
\eeq
as desired.\vspace{0.3cm}
\end{IEEEproof}

Define $\tilde \tau_2=\tau_1+\frac{-\log c}{D(f_{j'}||g_{j'})}$. Next we show that the sum LLRs of other cells are smaller than cell $j'$  at time $\tilde \tau_2$.\vspace{0.3cm}
\begin{lemma}
\label{lemma:other}
For every fixed $\epsilon>0$, there exists $C>0$ and $\gamma>0$, such that for all $j$ we have:
\beq
\mathbf {P}_m(S_{j'}(\tilde \tau_2)-S_j(\tilde \tau_2)<-\epsilon n)\le C e^{-\gamma n}, \hspace{0.5cm} \forall n>\tilde \tau_2. \vspace{0.2cm}
\eeq
\end{lemma}

\begin{IEEEproof}
For fixed $j$, define $t_0^j$ as the smallest integer such that $S_{j'}(n)<S_j(n)$ for all $t_0^j<i\le \tilde \tau_2$. By definition, $S_{j'}(t_0^j)\ge S_j(t_0^j)$. Then, by our selection rule, for all $t_0^j<i\le \tilde \tau_2$, whenever cell $j'$ is observed, cell $j$ must be observed based on their ranking of sum LLRs. Note that $D(f_{j'}||g_{j'})\le D(f_j||g_j)$. Thus, 
\beq
\bea{l}
\displaystyle\sum_{i=t_0^j}^{\tilde \tau_2} l_j(i)1_j(i)-\sum_{i=t_0^j}^{\tilde \tau_2} l_{j'}(i)1_{j'}(i) \\
\displaystyle=\sum_{i=t_0^j}^{\tilde \tau_2} \tilde l_j(i)1_j(i)-\sum_{i=t_0^j}^{\tilde \tau_2} \tilde l_{j'}(i)1_{j'}(i) \\
\displaystyle +D(f_j||g_j)\sum_{i=t_0^j}^{\tilde \tau_2} 1_j(i)-D(f_{j'}||g_{j'})\sum_{i=t_0^j}^{\tilde \tau_2} 1_{j'}(i) \\ 
\displaystyle \ge \sum_{i=t_0^j}^{\tilde \tau_2} \tilde l_j(i)1_j(i)-\sum_{i=t_0^j}^{\tilde \tau_2} \tilde l_{j'}(i)1_{j'}(i),
\ena
\eeq
which indicates that the LHS has positive means. By applying the generic Chernoff bound and using the i.i.d. property of $\tilde l_j(t_i)$ we have:
\beq
\mathbf {P}_m(S_{j'}(\tilde \tau_2)-S_{j'}(t_0^j)-(S_j(\tilde \tau_2)-S_j(t_0^j))<-\epsilon n)\le C e^{-\gamma n}, 
\eeq
for some $C,\gamma$. Since $S_{j'}(t_0^j)\ge S_j(t_0^j)$, we have:
\beq
\bea{l}
\displaystyle \mathbf {P}_m(S_{j'}(\tilde \tau_2)-S_j(\tilde \tau_2)<-\epsilon n) \vspace{0.2cm}\\
\displaystyle \le \mathbf {P}_m(S_{j'}(\tilde \tau_2)-S_{j'}(t_0^j)-(S_j(\tilde \tau_2)-S_j(t_0^j))<-\epsilon n) \vspace{0.2cm}\\
\displaystyle \le C e^{-\gamma n},\quad \forall n>\tilde \tau_2
\ena
\eeq
as desired. \vspace{0.3cm}
\end{IEEEproof}

Let $\tilde n_3\triangleq\tau_3-\tilde \tau_2$ denotes the total amount of time between $\tilde \tau_2$ and $\tau_3$. \vspace{0.2cm}

\begin{lemma}
\label{lem:tau3}
For every fixed $\epsilon>0$, there exists $C>0$ and $\gamma>0$ such that
\beq
\mathbf {P}_m(\tilde n_3>n)<Ce^{-\gamma n},\hspace{0.5cm} \forall n>-\epsilon \log c/D(f_{j'}||g_{j'}). \vspace{0.2cm}
\eeq
\end{lemma}

\begin{IEEEproof}
By substituting $t=\tilde \tau_2$ in Lemma~\ref{lemma:slow} we have:
\beq
\mathbf {P}_m(S_{j'}(\tilde \tau_2)>\log c+n\epsilon)<C_1 e^{-\gamma_1 n}  \hspace{0.5cm} \forall n>\tilde \tau_2
\eeq
for some $C_1,\gamma_1$. By applying Lemma~\ref{lemma:other}, we have:
\begin{equation}
\label{eq:other}
\bea{l}
\mathbf {P}_m(S_j(\tilde \tau_2)>\log c+n\epsilon)<C_2 e^{-\gamma_2 n}  \vspace{0.2cm}\\ \hspace{2.5cm} 
\forall n>\tilde \tau_2,j=1,2,\cdots,m
\ena
\end{equation}
for some $C_2,\gamma_2>0$.\\

Let $\tilde N_3^j$ denote that total number of observations, taken from cell $j$ between $\tilde \tau_2$ and $ \tau_3^j$. Since $\tilde n_3\le \sum \tilde N_3^j$, it suffices to show that $\mathbf {P}(\tilde N_3^j>n)$ decays exponentially with $n$. Note that
\beq
\bea{l}
\displaystyle \mathbf {P}_m(\tilde N_3^j>n)\vspace{0.2cm} \\ \hspace{-0.4cm}
\displaystyle \le \mathbf P_m\left(S_j(\tilde \tau_2)>\log c+n\frac{D(f_{j'}||g_{j'})}{2}\right) \vspace{0.2cm}\\
\displaystyle +\mathbf P_m\left(\tilde N_3^j>n|S_j(\tilde \tau_2)\le \log c+n\frac{D(f_{j'}||g_{j'})}{2}\right).
\ena
\eeq
By (\ref{eq:other}) it remains to show that the second term decays exponentially with $n$. Let $t_1,t_2,\cdots$ denote the time indices when cell $j$ is observed between $\tilde \tau_2$ and $\tau_3^j$. Then,
$$
\begin{aligned}
&\mathbf {P}_m(\tilde N_3^j>n|S_j(\tilde \tau_2) \le \log c+n\frac{D(f_{j'}||g_{j'})}{2}) \\
\le& \mathbf {P}_m\left(\inf_{r>n} \sum_{i=1}^r l_j(t_i)<n\frac{D(f_{j'}||g_{j'})}{2}\right) \\
\le& \mathbf {P}_m\left( \sum_{i=1}^r \tilde l_j(t_i)> r\frac{D(f_{j'}||g_{j'})}{2}\right).
\end{aligned}
$$
Applying the generic Chernoff bound and using the i.i.d.  property of $\tilde l_j(t_i)$ across time we have
\beq
\mathbf P_m\left( \sum_{i=1}^r \tilde l_j(t_i)> r\frac{D(f_{j'}||g_{j'})}{2}\right)<C_3 e^{-\gamma n}
\eeq
for some $C_3,\gamma_3$ which completes the proof.
\end{IEEEproof}

The following Lemma provides an upper bound on the detection time for the unbalanced case.

\begin{lemma}
\label{lemma:expected_time_policyu}
If DGFi policy is implemented, for the unbalanced case,  the expected detection time $\tau$ is upper bounded by:
\beq\label{eq:lemma_expected_time}
\E_m(\tau)\leq -\left(1+o(1)\right)\frac{\log(c)}{I_m (\Gamma_{\mbox{\footnotesize DGFi}})} \;,
\eeq
for $m=1, \ldots, M$.
\vspace{0.2cm}
\end{lemma}
%
\begin{IEEEproof}
Since the actual detection time under DGFi is upper bounded by: $\tau\leq\tau_3=\tilde \tau_2+\tilde n_3=\tau_1+\frac{-\log (c)}{I_m (\Gamma_{\mbox{\footnotesize DGFi}})}+\tilde n_3$, combining Lemmas~\ref{lemma:tau_1_policy1} and \ref{lem:tau3} proves the statement.
\vspace{0.2cm}
\end{IEEEproof}

Combining Lemma \ref{lemma:error_policy1} and Lemma \ref{lemma:expected_time_policyu}, Theorem 1 follows for the case of $K>1$.

\section*{Appendix B: Proof of Theorem~\ref{th:optimality_policy2}}
\label{appsub:lower}
First we show that in order to achieve a small order of Bayes Risk, $\Delta S_m(\tau)$ defined in~\eqref{eq:Delta_S_m} need to be sufficient large.

\begin{lemma}
\label{lemma:Delta_S_O}
Assume that $\alpha_j(\Gamma)=O(-c\log c)$ for all $j=1, \ldots, M$. Let $0<\epsilon<1$. Then:
\beq
\mathbf{P}_m\left(\Delta S_m(\tau)< -\left(1-\epsilon\right)\log c\;|\;\Gamma\right)=O(-c^{\epsilon}\log c) \;,
\eeq
for all $m=1, \ldots, M$.
\vspace{0.2cm} 
\end{lemma}
\begin{IEEEproof}
Note that:
\beq
\bea{l}
\mathbf{P}_m\left(\Delta S_m(\tau)< -\left(1-\epsilon\right)\log c|\Gamma\right) \vspace{0.2cm} \\
=\mathbf{P}_m\left(\Delta S_m(\tau)< -\left(1-\epsilon\right)\log c \;,\; \delta=m|\Gamma\right) \vspace{0.2cm} \\
+\mathbf{P}_m\left(\Delta S_m(\tau)< -\left(1-\epsilon\right)\log c \;,\; \delta\neq m|\Gamma\right) \vspace{0.2cm} \\
\leq\mathbf{P}_m\left(\Delta S_m(\tau)< -\left(1-\epsilon\right)\log c \;,\; \delta=m|\Gamma\right)+\alpha_m(\Gamma),
\ena
\eeq
where $\alpha_m(\Gamma)=O(-c\log c)$ by assumption. In what follows, we upper bound

$$
\mathbf{P}_m\left(\Delta S_m(\tau)< -\left(1-\epsilon\right)\log c \;,\; \delta=m|\Gamma\right).
$$

Similar to [2, Lemma 4] we can show that for all $j\neq m$ there exists $G>0$ such that:
\beq
\bea{l}
-Gc\log c \geq\mathbf{P}_j\left(\delta\neq j|\Gamma\right)\geq\mathbf{P}_j\left(\delta=m|\Gamma\right)  \vspace{0.2cm} \\
\geq \mathbf{P}_j\left(\Delta S_{m,j}(\tau)\leq -(1-\epsilon)\log c\;,\;\delta=m|\Gamma\right) \vspace{0.2cm} \\
\geq c^{1-\epsilon}\displaystyle\mathbf{P}_m\left(\Delta S_{m,j}(\tau)< -\left(1-\epsilon\right)\log c \;,\; \delta=m|\Gamma\right) \;,
\ena \vspace{0.2cm}
\eeq
where the last inequality holds by changing the measure as in [2, Lemma 4]. Thus,
\beq
\bea{l}
\displaystyle\mathbf{P}_m\left(\Delta S_{m,j}(\tau)< -\left(1-\epsilon\right)\log c \;,\; \delta=m|\Gamma\right)  \vspace{0.2cm} \\
\displaystyle=O\left(-c^{\epsilon}\log c\right) \;\;\;\forall j\neq m\;. \vspace{0.2cm}
\ena
\eeq
As a result, \vspace{0.2cm}
\beq
\bea{l}
\displaystyle\mathbf{P}_m\left(\Delta S_m(\tau)< -\left(1-\epsilon\right)\log c \;,\; \delta=m|\Gamma\right) \vspace{0.2cm}\\
\leq\displaystyle\sum_{j\neq m}\mathbf{P}_m\left(\Delta S_{m,j}(\tau)< -\left(1-\epsilon\right)\log c \;,\; \delta=m|\Gamma\right) 
 \vspace{0.2cm} \\
 \displaystyle=O\left(-c^{\epsilon}\log c\right) \;. \vspace{0.3cm}
\ena
\eeq
Finally, \vspace{0.2cm}
\beq
\bea{l}
\mathbf{P}_m\left(\Delta S_m(\tau)< -\left(1-\epsilon\right)\log c|\Gamma\right) =O\left(-c^{\epsilon}\log c\right).
\ena
\eeq
\vspace{0.2cm}
\end{IEEEproof}
%
\begin{lemma}
\label{lemma:inequality}
Assume that
\beq
D(g_m||f_m)\ge \frac{1}{\sum_{j\neq m}\frac {1}{D(f_j||g_j)}}.
\eeq
Then, the function:
\beq
\label{eq:func}
\bea{l}
\displaystyle d(t)\triangleq t\left[D(g_m||f_m)+\frac{\frac {n}{t}-1}{\sum_{j\neq m}\frac {1}{D(f_j||g_j)}}\right]  
\ena
\eeq
is monotonically increasing with $t$ for $0\le t\le n$.
\vspace{0.1cm}
\end{lemma}
\begin{IEEEproof}
Differentiation $d(t)$ with respect to $t$ yields:
\begin{center}
$\displaystyle \frac{\partial d(t)}{\partial t}=D(g_m||f_m)-\frac{1}{\sum_{j\neq m}\frac {1}{D(f_j||g_j)}}\geq 0 \;,$
\end{center}
which completes the proof. \vspace{0.3cm}
\end{IEEEproof}
%
For the next lemma we define
\beq
\label{eq:j_star}
j^*(t)\triangleq\arg\min_{j\neq m}N_j(t)D(f_j||g_j),
\eeq
and
\beq
\label{eq:W_star}
\bea{l}
\displaystyle W_m^*(t)\triangleq\sum_{i=1}^{t}\tilde{\ell}_m(i)\mathbf{1}_m(i)
-\sum_{i=1}^{t}\tilde{\ell}_{j^*(t)}(i)\mathbf{1}_{j^*(t)}(i) ,
\ena
\eeq
which is a sum of zero-mean random variable  \vspace{0.3cm}
\begin{lemma}
\label{lemma:W_star}
For every fixed $\epsilon>0$ there exist $C>0$ and $\gamma>0$ such that
\beq
\label{eq:lemma_W_star}
\bea{l}
\displaystyle \mathbf{P}_m\left(\max_{1\leq t\leq n}{W_m^*(t)}\geq n\epsilon|\Gamma\right)\leq Ce^{-\gamma n}
\ena
\eeq
for all $m=1, \ldots, M$ and for any policy $\Gamma$.
\vspace{0.1cm}
\end{lemma}
\begin{IEEEproof}
We upper bound (\ref{eq:lemma_W_star}) by summing over any possible values that $N_m(t), N_{j^*(t)}(t)$ can take and using the generic Chernoff bound given in~\eqref{gench}:
\beq
\label{eq:lemma_W_star_bound}
\bea{l}
\displaystyle \mathbf{P}_m\left(\max_{1\leq t\leq n}{W_m^*(t)}\geq n\epsilon|\Gamma\right) \vspace{0.2cm}\\
=\displaystyle \sum_{t=1}^{n}\;\sum_{i=0}^{t}\;\sum_{j=0}^{t}
\mathbf{P}_m\left(\sum_{r=1}^{t}\tilde{\ell}_m(r)\mathbf{1}_m(r) \right.\vspace{0.2cm}\\ 
\displaystyle \left.
                 +\sum_{r=1}^{t}-\tilde{\ell}_{j^*(t)}(r)\mathbf{1}_{j^*(t)}(r)\geq n\epsilon, 
                                                        N_m(t)=i,N_{j^*(t)}=j |\Gamma\right) \vspace{0.2cm}\\
\leq\displaystyle \sum_{t=1}^{n}\;\sum_{i=0}^{t}\;\sum_{j=0}^{t}
\displaystyle \left[\E_m\left(e^{s(\tilde{\ell}_m(1)-\epsilon/2)}\right)\right]^{i}
\vspace{0.2cm}\\ 
\displaystyle \times
\left[\E_m\left(e^{s(-\tilde{\ell}_{j^*(t)}(1)-\epsilon/2)}\right)\right]^{j}
\times
 \exp\left\{-s\frac{\epsilon}{2}(2n-i-j)\right\} \;,
\ena
\eeq
for all $s>0$.

Since $\E_m(\tilde{\ell}_m(1)-\epsilon/2)=-\epsilon/2<0$ and $\E_m(-\tilde{\ell}_{j^*(t)}(1)-\epsilon/2)=-\epsilon/2<0$ are strictly negative, using a similar argument as at the end of the proof of Lemma~\ref{lemma:S_j_S_m_N_j}, there exist $s>0$ and $\gamma'>0$ such that $\E_m\left(e^{s(\tilde{\ell}_m(1)-\epsilon/2)}\right)$, $\E_m\left(e^{s(-\tilde{\ell}_{j^*(t)}(1)-\epsilon/2)}\right)$ and $e^{-s\epsilon/2}$ are strictly less than $e^{-\gamma'}<1$. Since $2n-i-j\geq 0$, there exist $C>0$ and $\gamma>0$, such that summing over $t, i, j$ yields (\ref{eq:lemma_W_star}).
\vspace{0.3cm}
\end{IEEEproof}
%
%
%
\begin{lemma}
\label{lemma:Delta_S}
For any fixed $\epsilon>0$,
\beq
\label{eq:lemma_Delta_S}
\bea{l}
\displaystyle \mathbf{P}_m\left(\max_{1\leq t\leq n}{\Delta S_m(t)}\geq n\left(I_m^*+\epsilon\right)\;|\;\Gamma\right)\rightarrow 0 
         \;\; \mbox{as} \;\; n\rightarrow\infty \;,
\ena
\eeq
for all $m=1, \ldots, M$ and for any policy $\Gamma$.
\vspace{0.1cm}
\end{lemma}
\begin{IEEEproof}
We next show exponential decay of~\eqref{eq:lemma_Delta_S} (which is stronger than the polynomial decay shown under the binary composite hypothesis testing case in~\cite[Lemma $5$]{chernoff1959sequential}).
Let
\begin{center}
$\Delta S^*_m(t)\triangleq S_m(t)-S_{j^*(t)}(t)$.
\end{center}
Note that $\Delta S_m(t)\leq \Delta S^*_m(t)$ for all $m$ and $t$. As a result,
\beq
\label{eq:pr_lemma_Pr_Delta_S}
\bea{l}
\displaystyle \mathbf{P}_m\left(\max_{1\leq t\leq n}{\Delta S_m(t)}\geq n\left(I_m^*+\epsilon\right)|\Gamma\right)\vspace{0.2cm}\\
\leq\displaystyle\mathbf{P}_m\left(\max_{1\leq t\leq n}{\Delta S^*_m(t)}\geq n\left(I_m^*+\epsilon\right)|\Gamma\right).
\ena
\eeq

We next prove the lemma for the case where $I_m^*=F_m(K)$ and $u_m^*=0$. Proving the lemma for the cases where $u_m^*>0$ applies with minor modifications.

Note that:
\beq
\bea{l}
\displaystyle \Delta S^*_m(t)=W_m^*(t)+N_m(t)D(g_m||f_m)\vspace{0.2cm}\\ \hspace{1cm}
\displaystyle+N_{j^*(t)}(t)D(f_{j^*(t)}||g_{j^*(t)})
\vspace{0.2cm}\\ \hspace{1cm}
\displaystyle \leq W_m^*(t)+N_m(t)\cdot \frac {1}{\sum_{j\neq m}1/D(f_j||g_j)} \vspace{0.2cm}\\ 
\displaystyle +N_{j^*(t)}(t)D(f_{j^*(t)}||g_{j^*(t)}).
\ena
\eeq
Since that $j^*(t)=\arg\min_{j\neq m}N_j(t)D(f_j||g_j)$ and $Kt-N_m(t)$ is the total number of observations taken from $M-1$ cells $j\neq m$, we have:
\beq
\displaystyle \sum_{j\neq m} \frac{N_{j^*(t)}D(f_{j^*(t)}||g_{j^*(t)})}{D(f_j||g_j)}\le Kt-N_m(t)\le Kn-N_m(t).
\eeq
Hence,
\beq
\bea{l}
\displaystyle\Delta S^*_m(t)
\leq W_m^*(t)+ Kn \frac {1}{\sum_{j\neq m}1/D(f_j||g_j)}  \vspace{0.2cm}\\ \hspace{2cm}
=W_m^*(t)+n I_m^* \;.
\ena
\eeq
Therefore,
\begin{center}
$\Delta S^*_m(t)\geq n\left(I_m^*+\epsilon\right)$
\end{center}
implies
\begin{center}
$W_m^*(t)\geq n\epsilon$.
\end{center}
By Lemma~\ref{lemma:W_star} we have:
\beq
\bea{l}
\displaystyle \mathbf{P}_m\left(\max_{1\leq t\leq n}{\Delta S_m(t)}\geq n\left(I_m^*+\epsilon\right)\right) \\
\displaystyle \le \mathbf{P}_m\left(\max_{1\leq t\leq n}W_m^*(t)\geq n\epsilon\right) \\
\displaystyle \le Ce^{-\gamma n} \to 0 \mbox{ as } n\to\infty.
\ena
\eeq 
\vspace{0.3cm}
\end{IEEEproof}

Finally, we show that the Bayes risk cannot be made smaller than $\frac{-c\log(c)}{I_m^*}$:

\begin{lemma}
\label{th:lower_bound} 
Any policy $\Gamma$ that satisfies $R_j(\Gamma)=O(-c\log c)$ for all $j=1, \ldots, M$ must satisfy:
\beq
\label{eq:lower_bound}
\displaystyle R_m(\Gamma)\geq -\left(1+o(1)\right)\frac{c\log(c)}{I_m^*} \;.
\eeq
for all $m=1, \ldots, M$.
\vspace{0.2cm} \\
\end{lemma}
%
\begin{IEEEproof}
For any $\epsilon>0$ let $\displaystyle n_c=-(1-\epsilon)\frac{\log c}{I_m^*+\epsilon}$.
Note that
\beq
\bea{l}
\displaystyle\mathbf{P}_m\left(\tau\leq n_c\;|\;\Gamma\right) \vspace{0.2cm} \\
=\displaystyle\mathbf{P}_m\left(\tau\leq n_c \;,\; \Delta S_m(\tau)\geq -\left(1-\epsilon\right)\log c \;|\;\Gamma\right)   \vspace{0.2cm}\\
+\displaystyle\mathbf{P}_m\left(\tau\leq n_c \;,\; \Delta S_m(\tau)< -\left(1-\epsilon\right)\log c \;|\;\Gamma\right) \vspace{0.2cm} \\
\leq\displaystyle\mathbf{P}_m\left(\max_{t\leq n_c}\Delta S_m(t)\geq -\left(1-\epsilon\right)\log c \;|\;\Gamma\right)   \vspace{0.2cm}\\
+\displaystyle\mathbf{P}_m\left(\Delta S_m(\tau)< -\left(1-\epsilon\right)\log c \;|\;\Gamma\right) \vspace{0.2cm} .\\
\ena
\eeq
Both terms on the RHS approaches zero as $c\rightarrow 0$ by Lemmas~\ref{lemma:Delta_S_O},~\ref{lemma:Delta_S}. Hence,
\beq
\bea{l}
\displaystyle\E_m(\tau|\Gamma)\geq\sum_{n=n_c+1}^{\infty}n\mathbf{P}_m\left(\tau=n|\Gamma\right)
\vspace{0.2cm}\\\hspace{0.5cm}
\geq n_c\mathbf{P}_m\left(\tau\geq n_c+1|\Gamma\right)
\displaystyle\rightarrow n_c \mbox{\;\;\;as\;\;\;} c\rightarrow 0 \vspace{0.2cm}
\ena
\eeq
Since $\epsilon>0$ is arbitrarily small we have $\E_m(\tau|\Gamma)\geq-\left(1+o(1)\right)\log(c)/I_m^*$. As a result,  $R_m(\Gamma)\geq c\E_m(\tau|\Gamma)\geq -\left(1+o(1)\right)c\log(c)/I_m^*$.
\vspace{0.3cm}

\end{IEEEproof}

\section*{Appendix C: proof of Lemma~\ref{lemma:umstar}}
\label{prooflemma1}
Define
\beq
h_m(u)=uD(g_m||f_m)+F_m(K-u).
\eeq
By taking the derivative of $h_m(u)$, we have
\beq
h_m'(u)=D(g_m||f_m)-F_m'(K-u),
\eeq
where
\beq
F_m'(v)=\begin{cases}
\frac 1{\sum_{j\neq m}\frac {1}{D(f_j||g_j)}},  &\mbox{if } v\le \tilde K_m \\
\;0, &\mbox{if } v> \tilde K_m.
\end{cases}
\eeq

Since $F_m'(v)$ is piecewise constant with a breakpoint $\tilde K_m$, $h_m'(u)$  is piecewise constant with a breakpoint $K-\tilde K_m$. Therefore,
\begin{enumerate}
\item If  $D(g_m||f_m)\ge \bar F_m$, then $h_m'(u)>0$ and $u^*_m=1$. 
\item If $K>\tilde K_m+1$,  then $h_m'(u)=D(f_j||g_j)>0$  is a positive constant  and $u^*_m=1$
\item If  $D(g_m||f_m)<\bar F_m$ and $K<\tilde K_m$ then $h_m'(u)=D(g_m||f_m)<\bar F_m<0$ is a negative constant  and $u^*_m=0$
\item If none of the above is true, then $h_m'(u)>0$ for $u<K-\tilde K_m$ and $h_m'(u)>0$ for $u<K-\tilde K_m$. Therefore, $u^*_m=K-\tilde K_m$
\end{enumerate}

\section*{Appendix D: proof of Theorem~\ref{th:optimality_policyL}}
\label{thproofL}

We now focus on proving asymptotic optimality for $L>1$, and $K=1$. For $L>1$, we define $\tau_1$ as the smallest integer such that $S_m(n)> S_j(n)$ for all $m\in D$, $j\neq D$ and $n\geq\tau_1$. Note that when $K=1$ and $n\geq\tau_1$ the decision maker always probe the consistent cell (target or not depending on the order of $\bar G_D$ and $\bar F_D$) for making the difference between the $L^{th}$ and $(L+1)^{th}$ largest sum LLRs greater than the threshold $-\log c$. As a result, the decision maker can always balance the detection time so that the difference between the largest sum LLR
and the sum LLRs of any other cell exceeds the threshold $-\log c$ approximately at the same time as $c\rightarrow 0$. Thus, proving the asymptotic optimality of DGFi for $L>1$ and $K=1$ follows similar arguments as in the balanced case in the proof of Theorem 1 given in Appendix B, and we focus here only on the key modifications. Let
\beq
\label{eq:Delta_S_D}
\Delta S_{\mathcal D}(n)\triangleq\min_{m\in \mathcal D,j\notin \mathcal D} \Delta S_{m,j}(n),
\eeq
where $\Delta S_{m,j}(n)$ is defined in~\eqref{eq:Delta_S_m_j}. Without loss of generality we prove the theorem when set $\mathcal D$ contains all the targets. We define
\beq
\displaystyle \tilde{\ell}_k(i)=
\begin{cases} \ell_k(i)-D(g_k||f_k) \;,\;
                                                \mbox{if $k\in \mathcal D$,}   \vspace{0.3cm}\\
              \ell_k(i)+D(f_k||g_k) \;,\;
              \mbox{if $k\notin \mathcal D$,}
\end{cases}
\eeq
which is a zero-mean random variable. \vspace{0.2cm}

We start by showing the upper bound on the Bayes risk obtained by DGFi. Similar to Lemma 2, we can show that the error probability under DGFi is $O(c)$. Specifically, we can show that the error probability is upper bounded by:
\beq
\label{eq:error_policyL}
P_e\leq (M-L)L\cdot c \;.
\eeq
%

%
We can show this by letting $\alpha_{\mathcal D}=\mathbf P_{\mathcal D}(\delta\neq \mathcal D)$ and $\alpha_{\mathcal D,j}=\mathbf{P}_{\mathcal D}(j\in\delta)$ for all $j\notin\mathcal D$, where the subscript $D$ denotes the measure when set $D$ contains all the targets. Thus, $\alpha_{\mathcal D}\le \sum_{j\notin \mathcal D}\alpha_{\mathcal D,j}$.
By the stopping rule, accepting $j\in \delta$ implies $\Delta S_{j,m}\geq-\log c$ for some $m\in \mathcal D$.
Hence, for all $j\notin \mathcal D$ we have:
\beq
\bea{l}
\alpha_{\mathcal D,j}=\mathbf{P}_{\mathcal D}\left(j\in \mathcal D\right)\vspace{0.2cm} \\

\leq\sum_{m\in \mathcal D}\mathbf{P}_{\mathcal D}\left(\Delta S_{j,m}(\tau)\geq-\log c\right)  \vspace{0.2cm} \\
\leq \displaystyle\sum_{m\in \mathcal D}c\mathbf{P}_{\mathcal D\cup j\setminus m}\left(\Delta S_{j,m}(\tau)\geq -\log c\right)\leq L\cdot c,
\ena
\eeq
where we changed the measure in the second inequality. As a result,
\begin{center}
$\displaystyle\alpha_{\mathcal D}\le \sum_{j\notin \mathcal D}\alpha_{\mathcal D,j}\leq (M-L)L\cdot c$,
\end{center}
which yields~\eqref{eq:error_policyL}.
\vspace{0.3cm}

Here we consider the case where $I_{\mathcal D}=\bar G_{\mathcal D}$, the case $I_{\mathcal D}=\bar F_{\mathcal D}$ applies with minor modifications. For showing that $\tau_1$ is sufficiently small we need to show first the following Lemmas:

\begin{lemma}
\label{lemma:empty}
For all $j\notin D$, $\forall 0<q<1$, there exist $C,\gamma>0$ such that
\beq
\mathbf{P}_{\mathcal D}(N_j(n)>qn)<C e^{-\gamma n}
\eeq
\end{lemma}
\vspace{0.2cm}
\begin{IEEEproof}
For each $j$, define $t^j(n)$ as the time when cell $j$ is observed for the $n^{th}$ time. By DGFi selection rule, if cell $j$ is observed at time $t$, then there exists $m\in\mathcal D$ such that $S_j(t)\geq S_m(t)$. Hence,
\beq
\label{eq:164}
\bea{l}
\displaystyle \mathbf{P}_{\mathcal D}(N_j(n)>qn) \vspace{0.2cm}\\
\displaystyle \le \sum_{t=1}^n \mathbf{P}_{\mathcal D}(N_j(t)>qn, \exists m\in \mathcal D: S_j(t)>S_m(t) )  \vspace{0.2cm}\\ 
\displaystyle \times\mathbf{P}_{\mathcal D}(t^j(\lceil qn\rceil)=t).
\ena
\eeq

It suffices to show that there exist constants $C,\gamma$ such that
\beq
\label{eq:165}
\mathbf{P}_{\mathcal D}(N_j(t)>qn,\exists m\in \mathcal D: S_j(t)>S_m(t) ) \le C e^{-\gamma n}
\eeq
for all $t\le n$.

First we have 
\beq
\label{eq:166}
\bea{l}
\displaystyle \mathbf{P}_{\mathcal D}(N_j(t)>qn,\exists m\in \mathcal D: S_j(t)>S_m(t) ) \vspace{0.2cm}\\
\displaystyle \le \sum_{m\in \mathcal D} \mathbf{P}_{\mathcal D}(N_j(t)>qn, S_j(t)>S_m(t) ).
\ena
\eeq

Fix $m$, then we have
\beq
\bea{l}
\displaystyle \mathbf{P}_{\mathcal D}(N_j(t)>qn, S_j(t)>S_m(t) )\vspace{0.2cm}\\
\displaystyle \le \sum_{r=\lceil qn \rceil}^n \sum_{k=0}^n \mathbf{P}_{\mathcal D}\left (\sum_{i=1}^n\ell_j(i)+\sum_{k=1}^k-\ell_m(i)\ge 0\right)\vspace{0.2cm}\\
\displaystyle \le C_{m} e^{-\gamma_{m}n}.
\ena
\eeq
The last inequality can be shown using the generic Chernoff bound given in~\eqref{gench}.

To show~\eqref{eq:165}, we let $C=\sum_m C_m, \gamma=\min_m \gamma_m$, which completes the proof.
\end{IEEEproof}
\vspace{0.2cm}
\begin{lemma}
\label{lemma:target}
For all $m\in D$, and $\epsilon>0$, there exist $C,\gamma>0$ such that
\beq
\mathbf{P}_{\mathcal D}\left(N_m(n)>\frac{\bar G_{\mathcal D}}{D(g_m||f_m)-\epsilon}\cdot n\right)\le C e^{-\gamma n}
\eeq
\end{lemma}
\vspace{0.2cm}
\begin{IEEEproof}
For each $m$, define $t^m(n)$ as the time when cell $m$ is observed for the $n^{th}$ time. By DGFi selection rule, if cell $m$ is observed at time $t$, either there exists $j\notin D$ such that $S_j(n)>S_m(n)$ or $S_{m'}(n)>S_m(n)$ for all $m'\in D$. Similar to~\eqref{eq:164}, it suffices to show that
\beq
\label{eq:169}
\bea{l}
\mathbf{P}_{\mathcal D}\left(N_m(t)>\frac{\bar G_{\mathcal D}}{D(g_m||f_m)-\epsilon}\cdot n,\exists j\notin \mathcal D: S_j(t)>S_m(t)\right) 
\vspace{0.2cm}\\ 
\displaystyle 
 \le C e^{-\gamma n}
\ena
\eeq
and
\beq
\label{eq:170}
\bea{l}
\mathbf{P}_{\mathcal D}\left(N_m(t)>\frac{\bar G_{\mathcal D}}{D(g_m||f_m)-\epsilon}\cdot n,\forall m'\in \mathcal D: S_{m'}(t)>S_m(t)\right)
\vspace{0.2cm}\\ 
\displaystyle 
 \le C e^{-\gamma n}
\ena
\eeq
for all $t<n$.

\noindent Since~\eqref{eq:169} can be shown similarly as in~\eqref{eq:165}, it remains to show~\eqref{eq:170}.
By the definition of $\bar G_{\mathcal D}$, if $N_m(t)>\frac{\bar G_{\mathcal D}}{D(g_m||f_m)-\epsilon}\cdot n$, there exists $m'\in D$ and $\epsilon'>0$ such that $N_{m'}(t)<\frac{\bar G_{\mathcal D}}{D(g_m'||f_m')+\epsilon'}\cdot t$. Hence,
\beq
\label{eq:171}
\bea{l}
\displaystyle \mathbf{P}_{\mathcal D}\bigg(N_m(t)>\frac{\bar G_{\mathcal D}}{D(g_m||f_m)-\epsilon}\cdot n, \forall m'\in D: \vspace{0.2cm}\\ \hspace{2cm}
\displaystyle S_{m'}(t)>S_m(t) \bigg)\vspace{0.2cm}\\
\displaystyle\le\sum_{m'\in D}\mathbf{P}_{\mathcal D}\bigg(N_m(t)>\frac{\bar G_{\mathcal D}}{D(g_m||f_m)-\epsilon}\cdot n,S_{m'}(t)>S_m(t),\vspace{0.2cm}\\ 
\displaystyle N_m'(t)<\frac{\bar G_{\mathcal D}}{D(g_m'||f_m')+\epsilon'}\cdot t\bigg).
\ena
\eeq

\vspace{0.2cm}

Fix $m'$, and let $s_1=\frac{\bar G_{\mathcal D}}{D(g_m||f_m)-\epsilon}$, $s_2=\frac{\bar G_{\mathcal D}}{D(g_{m'}||f_{m'})+\epsilon}$. Then, we have
\beq
\bea{l}
\displaystyle\mathbf{P}_{\mathcal D}\bigg(N_m(t)>\frac{\bar G_{\mathcal D}}{D(g_m||f_m)-\epsilon}\cdot n, S_{m'}(t)>S_m(t)\vspace{0.2cm}\\ 
\displaystyle ,N_m'(t)<\frac{\bar G_{\mathcal D}}{D(g_{m'}||f_{m'})+\epsilon'}\cdot t\bigg) \vspace{0.2cm}\\
\displaystyle\leq\sum_{r=\lceil s_1 n\rceil}^{n}\;\sum_{k=0}^{\lfloor s_2t \rfloor}\mathbf{P}_{\mathcal D}\left(\sum_{i=1}^{r}-\ell_{m}(i)
                                                              +\sum_{i=1}^{k}\ell_{m'}(i)\geq 0\right)
\vspace{0.2cm}\\
\displaystyle \leq\sum_{r=\lceil s_1 n\rceil}^{n}\;\sum_{k=0}^{\lfloor s_2t \rfloor}\mathbf{P}_{\mathcal D}\left(\sum_{i=1}^{r}D(g_m||f_m)-\epsilon-\ell_{m}(i) \right.\vspace{0.2cm}\\ 
\displaystyle \left.      +\sum_{i=1}^{k}\ell_{m'}(i)-D(g_{m'}||f_{m'})-\epsilon'\geq 0\right)
\vspace{0.2cm}\\

\displaystyle\leq\sum_{r=\lceil s_1 n\rceil}^{n}\;\sum_{k=0}^{\lfloor s_2t \rfloor}\left[\E_{\mathcal D}\left(e^{s (-\tilde\ell_m(1)-\epsilon)}\right)\right]^{r}
    \left[\E_{\mathcal D}\left(e^{s(\ell_{m'}(1)-\epsilon')}\right)\right]^{k} \vspace{0.2cm}\\
\displaystyle\leq C_{m'}e^{-\gamma_{m'} n}
\ena
\eeq
The last inequality can be shown using the generic Chernoff bound given in~\eqref{gench}. To show~\eqref{eq:171}, we let $C=\sum_{m'} C_{m'}, \gamma=\min_{m'} \gamma_{m'}$, which completes the proof.
\vspace{0.2cm}
\end{IEEEproof}

\begin{lemma}
\label{targetfreq}
For all $m\in D$, $\forall \epsilon>0$, there exist $C,\gamma>0$ such that
\beq
\mathbf{P}_{\mathcal D}\left(N_m(n)<(\frac{\bar G_{\mathcal D}}{2D(g_m||f_m)})n\right)\le C e^{-\gamma n}.
\eeq
\end{lemma}
\vspace{0.2cm}
\begin{IEEEproof}
By choosing $q_j$ and $\epsilon_m'$ in Lemma~\ref{lemma:empty} and Lemma~\ref{lemma:target} such that $\sum_j q_j+\sum_m' \epsilon_m'=\frac{\bar \bar G_{\mathcal D}}{2D(g_m||f_m)}$, we have
\beq
\bea{l}
\displaystyle\mathbf{P}_{\mathcal D}\left(N_m(n)<(\frac{\bar G_{\mathcal D}}{2D(g_m||f_m)})n\right)\\
\displaystyle\le\sum_{j\notin \mathcal D} \mathbf{P}_{\mathcal D}\left(N_j(n)>q_jn\right)\\ \hspace{0.0cm}
\displaystyle+\sum_{m'\in \mathcal D} \mathbf{P}_{\mathcal D}\left(N_m'(n)>(\frac{\bar G_{\mathcal D}}{D(g_m'||f_m')}+\epsilon_m')n\right)\le C_{m'} e^{-\gamma n}
\ena
\eeq
as desired.
\end{IEEEproof}
\vspace{0.2cm}

Next, similar to Lemma 4, we can show that the probability that $\tau_1$ is greater than $n$ decreases exponentially with $n$. This result is used when evaluating the asymptotic expected search time to show that it is not affected by $\tau_1$. We can show this by noting that
\beq\label{eq:tau1L}
\bea{l}
\displaystyle\mathbf{P}_{\mathcal D}\left(\tau_1>n\right)\leq\mathbf{P}_{\mathcal D}\left(\max_{j\notin \mathcal D,m\in \mathcal D}\;\sup_{t\geq n}\;\left(S_j(t)-S_m(t)\right)\geq 0 \right) \vspace{0.2cm} \\ \hspace{2cm}
\leq\displaystyle\sum_{j\notin \mathcal D, m\in \mathcal D}\;\sum_{t=n}^{\infty}\mathbf{P}_{\mathcal D}\left(S_j(t)\geq S_m(t)\right)\;.
\ena
\eeq
Following~\eqref{eq:tau1L}, it suffices to show that $\mathbf{P}_{\mathcal D}\left(S_j(n)\geq S_m(n)\right)$ decays exponentially with $n$. Note that
\beq
\bea{l}
\displaystyle\mathbf{P}_{\mathcal D}\left(S_j(n)\geq S_m(n)\right)\\
\displaystyle\leq\mathbf{P}_{\mathcal D}\left(S_j(n)\geq S_m(n),N_m(n)\ge(\frac{\bar G_{\mathcal D}}{2D(g_m||f_m)})n\right)\\ \hspace{0.5cm}
\displaystyle+\mathbf{P}_{\mathcal D}\left(N_m(n)<(\frac{\bar G_{\mathcal D}}{2D(g_m||f_m)})n\right)
\ena
\eeq
The first term decays exponentially with $n$ by Lemma~\ref{lemma:S_j_S_m_N_j} (with minor modifications). The second term decays exponentially with $n$ by Lemma~\ref{targetfreq}.

Note that we obtained that the expectation of $\tau_1$ is bounded, and we can use similar arguments as in the balanced case of Theorem 1 in Appendix B to obtain the detection rate $I_D$ for $n\geq\tau_1$. Combining these results yields that the expected detection time $\tau$ under the DGFi policy is upper bounded by:
\beq\label{eq:lemma_expected_timeL}
\E_{\mathcal D}(\tau)\leq -\left(1+o(1)\right)\frac{\log(c)}{I_{\mathcal D}} \;,
\eeq
for $m=1, \ldots, M$.
\vspace{0.2cm}

Finally, showing that the asymptotic Bayes risk is lower bounded by $-c\log c/I^*_L$ follows a similar outline as in Appendix B. Specifically, similar to Lemma~\ref{lemma:Delta_S_O}, if $\alpha_{\mathcal D}(\Gamma)=O(-c\log c)$ for all $\mathcal D$, and we let $0<\epsilon<1$, then:
\beq
\mathbf{P}_{\mathcal D}\left(\Delta S_m(\tau)< -\left(1-\epsilon\right)\log c\;|\;\Gamma\right)=O(-c^{\epsilon}\log c) \;,
\eeq
for all $\mathcal D$ and $m\in \mathcal D$.
\vspace{0.2cm} 
Then, we define:
\beq
\label{eq:j_starL}
j^*(t)\triangleq\arg\min_{j\notin \mathcal D}N_j(t)D(f_j||g_j),
\eeq
\beq
\label{eq:m_star}
m^*(t)\triangleq\arg\min_{m\in \mathcal D}N_{m^*(t)}(t)D(g_m||f_m),
\eeq

and
\beq
\label{eq:W_starL}
\bea{l}
\displaystyle W_{\mathcal D}^*(t)\triangleq\sum_{i=1}^{t}\tilde{\ell}_{m^*(t)}(i)\mathbf{1}_{m^*(t)}(i)
-\sum_{i=1}^{t}\tilde{\ell}_{j^*(t)}(i)\mathbf{1}_{j^*(t)}(i),
\ena
\eeq
where $W_{\mathcal D}^*(t)$ is a sum of zero-mean random variable. Using these definitions, similar to Lemma~\ref{lemma:W_star}, we can show that for every fixed $\epsilon>0$ there exist $C>0$ and $\gamma>0$ such that
\beq
\label{eq:lemma_W_starL}
\bea{l}
\displaystyle \mathbf{P}_{\mathcal D}\left(\max_{1\leq t\leq n}{W_{\mathcal D}^*(t)}\geq n\epsilon|\Gamma\right)\leq Ce^{-\gamma n}
\ena
\eeq
for all $\mathcal D$ and for any policy $\Gamma$.
\vspace{0.1cm}

Next, similar to Lemma~\ref{lemma:Delta_S} we can show that for any fixed $\epsilon>0$, 
\beq
\label{eq:lemma_Delta_SL}
\bea{l}
\displaystyle \mathbf{P}_{\mathcal D}\left(\max_{1\leq t\leq n}{\Delta S_{\mathcal D}(t)}\geq n\left(I_{\mathcal D}+\epsilon\right)\;|\;\Gamma\right)\rightarrow 0 \vspace{0.2cm} \\ \hspace{6cm}
         \;\; \mbox{as} \;\; n\rightarrow\infty \;,
\ena
\eeq
for all $\mathcal D$ and for any policy $\Gamma$.

Finally, similar to Lemma~\ref{th:lower_bound}, we can show that any policy $\Gamma$ that satisfies $R_{\mathcal D}(\Gamma)=O(-c\log c)$ for all $\mathcal D$ must satisfy:
\beq
\label{eq:lower_boundL}
\displaystyle R_{\mathcal D}(\Gamma)\geq -\left(1+o(1)\right)\frac{c\log(c)}{I_{\mathcal D}} \;.
\eeq
for all $\mathcal D$.
\vspace{0.2cm} \\
%

\bibliographystyle{ieeetr}
\bibliography{DGFi_IT_Final}

\begin{thebibliography}{10}

\bibitem{zhao2007survey}
Q.~Zhao and B.~M. Sadler, ``A survey of dynamic spectrum access,'' {\em IEEE
  signal processing magazine}, vol.~24, no.~3, pp.~79--89, 2007.

\bibitem{chernoff1959sequential}
H.~Chernoff, ``Sequential design of experiments,'' {\em The Annals of
  Mathematical Statistics}, vol.~30, no.~3, pp.~755--770, 1959.

\bibitem{wald1947sequential}
A.~Wald, ``Sequential analysis. 1947,'' {\em Zbl0029}, vol.~15805, 1947.

\bibitem{bessler1960theory}
S.~A. Bessler, ``Theory and applications of the sequential design of
  experiments, k-actions and infinitely many experiments. part i. theory,''
  tech. rep., DTIC Document, 1960.

\bibitem{nitinawarat2012controlled}
S.~Nitinawarat, G.~K. Atia, and V.~V. Veeravalli, ``Controlled sensing for
  hypothesis testing,'' in {\em 2012 IEEE International Conference on
  Acoustics, Speech and Signal Processing (ICASSP)}, pp.~5277--5280, IEEE,
  2012.

\bibitem{nitinawarat2013controlled}
S.~Nitinawarat, G.~K. Atia, and V.~V. Veeravalli, ``Controlled sensing for
  multihypothesis testing,'' {\em IEEE Transactions on Automatic Control},
  vol.~58, no.~10, pp.~2451--2464, 2013.

\bibitem{nitinawarat2015controlled}
S.~Nitinawarat and V.~V. Veeravalli, ``Controlled sensing for sequential
  multihypothesis testing with controlled markovian observations and
  non-uniform control cost,'' {\em Sequential Analysis}, vol.~34, no.~1,
  pp.~1--24, 2015.

\bibitem{naghshvar2013active}
M.~Naghshvar and T.~Javidi, ``Active sequential hypothesis testing,'' {\em The
  Annals of Statistics}, vol.~41, no.~6, pp.~2703--2738, 2013.

\bibitem{naghshvar2013sequentiality}
M.~Naghshvar and T.~Javidi, ``Sequentiality and adaptivity gains in active
  hypothesis testing,'' {\em IEEE Journal of Selected Topics in Signal
  Processing}, vol.~7, no.~5, pp.~768--782, 2013.

\bibitem{castanon1995optimal}
D.~A. Castanon, ``Optimal search strategies in dynamic hypothesis testing,''
  {\em IEEE transactions on systems, man, and cybernetics}, vol.~25, no.~7,
  pp.~1130--1138, 1995.

\bibitem{cohen2014optimal}
K.~Cohen, Q.~Zhao, and A.~Swami, ``Optimal index policies for anomaly
  localization in resource-constrained cyber systems,'' {\em IEEE Transactions
  on Signal Processing}, vol.~62, no.~16, pp.~4224--4236, 2014.

\bibitem{vaidhiyan2015learning}
N.~K. Vaidhiyan and R.~Sundaresan, ``Learning to detect an oddball target,''
  {\em arXiv preprint arXiv:1508.05572}, 2015.

\bibitem{leahy2016always}
K.~Leahy and M.~Schwager, ``Always choose second best: Tracking a moving target
  on a graph with a noisy binary sensor,'' in {\em Control Conference (ECC),
  2016 European}, pp.~1715--1721, IEEE, 2016.

\bibitem{heydari2016quickest}
J.~Heydari, A.~Tajer, and H.~V. Poor, ``Quickest linear search over correlated
  sequences,'' {\em IEEE Transactions on Information Theory}, vol.~62, no.~10,
  pp.~5786--5808, 2016.

\bibitem{cohen2015active}
K.~Cohen and Q.~Zhao, ``Active hypothesis testing for anomaly detection,'' {\em
  IEEE Transactions on Information Theory}, vol.~61, no.~3, pp.~1432--1450,
  2015.

\bibitem{zigangirov1966problem}
K.~S. Zigangirov, ``On a problem in optimal scanning,'' {\em Theory of
  Probability \& Its Applications}, vol.~11, no.~2, pp.~294--298, 1966.

\bibitem{tajer2013quick}
A.~Tajer and H.~V. Poor, ``Quick search for rare events,'' {\em IEEE
  Transactions on Information Theory}, vol.~59, no.~7, pp.~4462--4481, 2013.

\bibitem{cohen2015asymptotically}
K.~Cohen and Q.~Zhao, ``Asymptotically optimal anomaly detection via sequential
  testing,'' {\em IEEE Transactions on Signal Processing}, vol.~63, no.~11,
  pp.~2929--2941, 2015.

\bibitem{fellouris2017multistream}
G.~Fellouris, G.~V. Moustakides, and V.~V. Veeravalli, ``Multistream quickest
  change detection: Asymptotic optimality under a sparse signal,'' in {\em
  Acoustics, Speech and Signal Processing (ICASSP), 2017 IEEE International
  Conference on}, pp.~6444--6447, IEEE, 2017.

\bibitem{lai2011quickest}
L.~Lai, H.~V. Poor, Y.~Xin, and G.~Georgiadis, ``Quickest search over multiple
  sequences,'' {\em IEEE Transactions on Information Theory}, vol.~57, no.~8,
  pp.~5375--5386, 2011.

\bibitem{nitinawarat2015universal}
S.~Nitinawarat and V.~V. Veeravalli, ``Universal scheme for optimal search and
  stop,'' in {\em Information Theory and Applications Workshop (ITA), 2015},
  pp.~322--328, IEEE, 2015.

\bibitem{hemo2016}
B.~Hemo, K.~Cohen, and Q.~Zhao, ``Asymptotically optimal search of unknown
  anomalies,'' in {\em Proc. of the 16th IEEE Symposium on Signal Processing
  and Information Technology (ISSPIT)}, (Limassol, Cyprus), Dec. 2016.

\bibitem{malloy2012quickest}
M.~L. Malloy, G.~Tang, and R.~D. Nowak, ``Quickest search for a rare
  distribution,'' in {\em Information Sciences and Systems (CISS), 2012 46th
  Annual Conference on}, pp.~1--6, IEEE, 2012.

\bibitem{pei2011energy}
Y.~Pei, Y.-C. Liang, K.~C. Teh, and K.~H. Li, ``Energy-efficient design of
  sequential channel sensing in cognitive radio networks: optimal sensing
  strategy, power allocation, and sensing order,'' {\em IEEE Journal on
  Selected Areas in Communications}, vol.~29, no.~8, pp.~1648--1659, 2011.

\bibitem{caromi2013fast}
R.~Caromi, Y.~Xin, and L.~Lai, ``Fast multiband spectrum scanning for cognitive
  radio systems,'' {\em IEEE Transactions on Communications}, vol.~61, no.~1,
  pp.~63--75, 2013.

\bibitem{ferrari2017utility}
L.~Ferrari, Q.~Zhao, and A.~Scaglione, ``Utility maximizing sequential sensing
  over a finite horizon,'' {\em IEEE Transactions on Signal Processing},
  vol.~65, no.~13, pp.~3430--3445, 2017.

\bibitem{egan2017fast}
M.~Egan, J.-M. Gorce, and L.~Cardoso, ``Fast initialization of cognitive radio
  systems,'' in {\em IEEE International Workshop on Signal Processing Advances
  in Wireless Communications}, 2017.

\bibitem{tajer2014outlying}
A.~Tajer, V.~V. Veeravalli, and H.~V. Poor, ``Outlying sequence detection in
  large data sets: A data-driven approach,'' {\em IEEE Signal Processing
  Magazine}, vol.~31, no.~5, pp.~44--56, 2014.

\bibitem{chandola2009anomaly}
V.~Chandola, A.~Banerjee, and V.~Kumar, ``Anomaly detection: A survey,'' {\em
  ACM computing surveys (CSUR)}, vol.~41, no.~3, p.~15, 2009.

\bibitem{bhuyan2014network}
M.~H. Bhuyan, D.~K. Bhattacharyya, and J.~K. Kalita, ``Network anomaly
  detection: methods, systems and tools,'' {\em IEEE Communications Surveys \&
  Tutorials}, vol.~16, no.~1, pp.~303--336, 2014.

\bibitem{huang2017sequential}
B.~Huang, K.~Cohen, and Q.~Zhao, ``Sequential active detection of anomalies in
  heterogeneous processes,'' {\em arXiv preprint arXiv:1704.00766}, 2017.

\end{thebibliography}

\end{document}